\documentclass[12pt]{article}
\usepackage[dvips]{graphicx}
\usepackage{amsbsy}
\usepackage{amssymb}
\usepackage{float}
\usepackage{wrapfig}
\usepackage{color}

\newcommand{\beq}{\begin{equation}}
\newcommand{\eeq}{\end{equation}}
\newcommand{\bdm}{\begin{displaymath}}
\newcommand{\edm}{\end{displaymath}}
\newcommand{\beqr}{\begin{eqnarray}}
\newcommand{\eeqr}{\end{eqnarray}}
\newcommand{\beqrn}{\begin{eqnarray*}}
\newcommand{\eeqrn}{\end{eqnarray*}}

\def\sech{{\rm sech}}
\begin{document}

\title{{\bf On self-gravitating kinks in two-dimensional pseudo-riemannian universes}}

\author{A. Alonso Izquierdo$^\dag$, W. Garc\'{\i}a Fuertes$^*$,\\
J. Mateos Guilarte$^{\dag\dag}$}
\date{}
\maketitle
\begin{center}
$^{\dag}${{\it Departamento de Matem\'atica Aplicada, Facultad de Ciencias Agrarias y Ambientales,  Universidad de Salamanca,
E-37008 Salamanca, Spain.}}
\\ \vspace{0.2cm} $^*${{\it Departamento de F\'{\i}sica, Facultad de Ciencias,  Universidad de Oviedo,\\
E-33007 Oviedo, Spain.}}
\\ \vspace{0.2cm}
$^{\dag\dag}${{\it IUFFyM, Facultad de Ciencias,  Universidad de Salamanca,
E-37008 Salamanca, Spain.}}
\vskip1cm
\end{center}

\date{ }

\maketitle

\begin{abstract}
\noindent
Two-dimensional scalar field theories with spontaneous symmetry breaking subject to the action of Jackiw-Teitelboim gravity are studied. Solutions for the $\phi^4$ and sine-Gordon self-gravitating kinks are presented, both for general gravitational coupling and in the perturbative regime. The analysis is extended to deal with a hierarchy of kinks related to transparent P\"{o}schl-Teller potentials. 
\end{abstract}



\vfill\eject 
\section{Introduction}

At the crossroad of general relativity, elementary particle theory, phase transitions in the early Universe and Cosmology, emerged in the early eighties of the past Century many interesting developments where topological defects were studied in connection with gravitational fields, see e.g. Kibble and Vilenkin, \cite{Kibble,VilShel,Vachaspati2006}. At first, the attention was mainly devoted to consider the topological defects as sources of gravitational fields, and later the backreaction of the gravitational field on the source was also investigated, although this is a much more difficult problem. As an attempt to address both issues in the clearest possible terms, we choose here to work in low dimensional universes with low dimensional kinks. Observational constraints on the evolution of the cosmic microwave background (CMB) seems to give importance mainly to strings, either primordial or solitonic. For example, convolutional neural network are used in recent references, see \cite{Ciuca2020}, to study the cosmic string formation pattern in the CMB. Large-scale simulations of an expanding-universe string network are analyzed in \cite{JJ2015}. On the theoretical side, different types of theories containing strings can be considered and, in particular, the sigma model versions of the Abelian Higgs model studied in \cite{AGG15} offer a rich phenomenology. We postpone for a later investigation the analysis of self-gravitating cosmic strings of this type in  $(2+1)$-pseudo-Riemannian Universes. In the mean-time, we address this issue in linear gravity coupled to a real scalar field on a line.

As it is well known, general relativity in lower dimensions is quite peculiar \cite{Br88, Co77}. In three dimensions, Einstein equations give a viable theory, albeit spacetime turns out to be flat around the sources and these do not create a gravitational field, only conical singularities \cite{DJtH84}. The situation is worse in 1+1 dimensions, where the Einstein-Hilbert action, being a topological invariant, the Euler class, has identically vanishing  variations with respect to the metric. Hence, two-dimensional Einstein equations only make sense if the energy-momentum tensor of matter is zero everywhere. A sensible theory of gravity in 1+1 dimensions requires thus a different dynamics. Given that in two dimensions the only independent component of the Riemann curvature tensor is directly determined by the curvature scalar $R$, the simplest vacuum gravitational equation in sight is 
\beq
R=\Lambda\label{eq:jtvac}
\eeq
where $\Lambda$ is a cosmological constant. Despite its simplicity, this equation can be derived from a variational principle only at the price of putting aside some of the tenets of Einstein general relativity. Thus, Jackiw \cite{Ja8485} found an action for (\ref{eq:jtvac}) which, although generally covariant, sacrifices the equivalence principle by depending not only on the metric but also on an auxiliary Lagrange multiplier field. Alternatively, Teitelboim \cite{Te8384} was able to find an action leading to (\ref{eq:jtvac}) which depends only on the metric, but this time  giving up general covariance. In any case, Jackiw-Teitelboim gravity is a very remarkable theory with many interesting properties, see \cite{Br88} for review. It is, for instance, directly related to bosonic string theory. This comes about because, although the classical Einstein-Hilbert action is trivial in two dimensions, when the path integral is performed in conformal gauge $g_{\mu\nu}=\eta_{\mu\nu} e^\phi$ there is a contribution from the conformal anomaly. This generates a quantum field theory for $\phi$ with Liouville action \cite{Po81}, but it turns out that, in this gauge,  the Liouville equation is exactly (\ref{eq:jtvac}) \cite{Ja8485}. Other noteworthy aspect of JT gravity is that it is equivalent to a gauge field theory with gauge group $SO(2,1)$: if $P_0, P_1$ and $J$ are the generators of the Lie algebra $so(2,1)$, one defines this theory by introducing a gauge field $A_\mu=e^a_\mu P_a+\omega_\mu J$, with $e_\mu^a$ the zweibein and $\omega_\mu$ the spin connection, and using a gauge invariant Lagrangian ${\cal L}=\eta_a F^a$ where $\eta_a$ are auxiliary fields in the co-adjoint representation. Once the Euler-Lagrange equations for these field are taken into account, the action of the gauge theory coincides with the action for (\ref{eq:jtvac}) proposed by Jackiw, see\cite{FuKa87,Ja92}. 

On the other hand, (\ref{eq:jtvac}) can be generalized to take care of the presence of matter. This  leads to the equation
\beq
R-\Lambda=8\pi G T\label{eq:jtmat},
\eeq
where $T$ is the trace of the energy-momentum tensor, this new theory being also susceptible of being derived from a local action \cite{To88,MaMoSiSt91}. It is to be noted that (\ref{eq:jtmat}) is the trace of Einstein equations in four dimensions (although with the wrong sign for the coupling) and, in fact, it can be argued in various ways that JT with matter is the closest relative to Einstein theory in 1+1 dimensions. For instance, conventional general relativity  in higher dimensions can be thought of as a limit version of Brans-Dicke theory in which the kinetic term of the scalar field is multiplied by a constant which approaches infinity. As it has been shown in \cite{LeSa94}, performing the same limit in 1+1 dimensions leads to JT theory with sources. Another argument is that the action for (\ref{eq:jtmat}) can be recovered from the Einstein-Hilbert action in $D$ dimensions by taking the limit $D\rightarrow 2$ if, at the same time, the gravitational coupling $G_D$ is made to decrease at the same pace than the Einstein tensor does, see \cite{MaRo93}. Apart of its relation to general relativity, JT gravity is also interesting because it is one of the simplest examples of the so-called dilaton gravity theories in two dimensions, see \cite{GrKuVa02} for a comprehensive review, and furthermore it is singular among these in that, as we have said, it admits a Lie algebra valued gauge version, while the gauge realization of more general dilatonic theories requires the use of central extensions of Lie algebras or W-algebras \cite{GrKuVa02,IkIz9394}. At any rate, the gravity theory of Jackiw and Teitelboim has attracted considerable attention and its phenomenology has been widely studied over the years, especially from the point of view of black hole physics, but also in other contexts such as stellar structure, gravitational waves, cosmology, etc; some references are \cite{BrHeTe86,MaShTa90,MaMoSiSt91,SiMa91,MoMa91,CaMi02}
and other listed in \cite{GrKuVa02}.

Scalar field theories display also some special features when they are formulated in two dimensions. If the potential has degenerate minima, these theories can harbor topological kinks, stable static solutions of the Euler-Lagrange equations whose energy is localized and finite, and which interpolate between different classical vacua. Due to Derrick's theorem \cite{De64}, for a relativistic theory with the canonical form of the Lagrangian, this is not possible in higher dimensions, where the existence of regular solitons requires the presence of gauge fields (exceptions occur, however, if the potential is allowed to include Lorentz invariant terms depending on the coordinates \cite{BaMeMe03}). Thus, as it was the case for JT gravity, topological kink solutions are another characteristic two-dimensional paradigm, one that has been intensely studied both by taking the kinks as classical field configurations as well as treating them as quantum particles, see \cite{Ra82} for a pedagogical exposition, or \cite{nuestro1,nuestro2} and references therein for more recent results on the quantum aspect of kinks. Also, kink solutions have been found useful for a variety of physical applications in fields like cosmology, supersymmetric gauge theory or solid state physics, see for instance \cite{BaInLo04}, which includes a collection of references about these topics. 

It is thus interesting to merge these two facets of two-dimensional physics and investigate the coupling of JT gravity to scalar fields theories allowing for the presence of kinks. In principle, a phenomenon to be considered is the formation of a black hole due to gravitational collapse of a lump of scalar matter which is subject to topologically non-trivial boundary conditions at infinity. Nevertheless, it seems that a kind of no-hair theorem operates here: once the black hole forms, the scalar field is forced by gravitational attraction to settle to a classical vacua out of the horizon, and we finish with a profile in which the field takes two different constant values, corresponding to minima of the potential, at each side of the black hole. However, it is clear that for field configurations sitting asymptotically onto two different vacua, the gravitational pull towards the center of the lump produces also an increase of the gradient potential energy of the scalar matter. This suggests that there should be some static solutions in which the gravitational attraction and the repulsive gradient energy are in equilibrium and a self-gravitating kink with a non-trivial profile forms. The purpose of this paper is to study several static configurations of this type. This is in contrast with former research in a similar scenario, see References \cite{Stoet} and \cite{Gegen}, where the general structure of two-dimensional dilatonic gravity coupled to the sine-Gordon scalar field model is discussed.

The organization of the paper is as follows. In Section 2 we describe the general setting and examine the interaction between JT gravity and the most prototypical system allowing for topological kinks, the $\phi^4$ theory with broken discrete ${\mathbb Z}_2$-symmetry, both for general gravitational coupling and in the weak coupling regime. Then, we present in Section 3 the analysis of another important case, this time with a non-polynomial potential, the sine-Gordon theory. The treatment given for the $\phi^4$ and sine-Gordon kinks reveals that the underlying supersymmetry present in both models has an important role in the procedure to find the gravitating solutions. Thus, in Section 4 we extend the approach to deal with general kinks related to unbroken supersymmetry and apply it to a hierarchy of scalar field theories of this type,
which includes the  $\phi^4$ and sine-Gordon ones as the two first members. Finally, we offer in Section 5 some concluding remarks.
\section{Self-gravitating kinks in the $\phi^4$ theory}
\subsection{Scalar kink configurations coupled to Jackiw-Teitelboim gravity}
The purpose in this paper is to describe some classical static solutions which arise within the theory of a 1+1 dimensional real scalar field $\phi(t,x)$ when the dynamics is governed not only by the presence of a non-linear potential allowing for topologically non-trivial boundary conditions, but also by the existence of gravitational forces of the Jackiw-Teitelboim type. It turns out that, in order to define the action of such a system, one needs to introduce, apart of the $\phi(t,x)$ field itself and the metric tensor $g_{\mu\nu}(t,x)$, another auxiliary scalar field \cite{MaMoSiSt91}. This extra field can take the form of a mere Lagrange multiplier $N(t,x)$ or, alternatively, of a field $\Psi(t,x)$ which couples to the curvature scalar in a way which closely resembles the analogous coupling of the dilatonic field in string theory. Here, we will choose this second possibility. Thus, the action is
\bdm
S=\frac{1}{16\pi G}\int d^2x \sqrt{-g}\left( \Psi R+\frac{1}{2} g^{\mu\nu}\partial_\mu\Psi\partial_\nu\Psi+\Lambda\right)+S_M
\edm
where  $\Psi=e^{-2 \Phi}$ with $\Phi$ the dilaton field, and the matter action is
\bdm
S_M=-\int d^2x \sqrt{-g}\left(\frac{1}{2}g^{\mu\nu}\partial_\mu\phi\partial_\nu\phi+V(\phi)\right).
\edm
Upon variation of $S$, one arrives to two field equations
\beqr
\nabla_\mu \nabla^\mu \phi&=&\frac{dV}{d\phi}\label{eqjt1}\\
R-\Lambda&=&8\pi G T\label{eqjt2}
\eeqr
in which $\Psi$ decouples. Here $\nabla_\mu$ is the standard covariant derivative and $T$ is the trace of the energy-momentum tensor for the scalar field
\bdm
T_{\mu\nu}=\partial_\mu\phi\partial_\nu\phi-g_{\mu\nu}\left(\frac{1}{2} \partial_\sigma\phi\partial^\sigma\phi+V(\phi)\right).
\edm
With suitable boundary conditions, these equations are enough to completely determine the field $\phi$ and the metric. Once these are known, the auxiliary field is obtained from the remaining Euler-Lagrange equation
\bdm
\nabla_\mu \nabla^\mu \Psi=R .
\edm
In 1+1 dimensions the metric has three independent components, which are subject to arbitrary reparametrizations of the two coordinates. Thus, the metric can be put locally in a form which depends only on a single function of the coordinates. We will choose a gauge which is commonly used in problems related to black holes, see for instance \cite{MaMoSiSt91,MaShTa90}, although in our case the metric will be regular
\bdm
ds^2=-\alpha(t,x) dt^2+\frac{1}{\alpha(t,x)} dx^2.
\edm
In this gauge, the Christoffel symbols are 
\bdm
\Gamma^0_{00}=-\Gamma^1_{01}=\frac{1}{2\alpha}\frac{\partial\alpha}{\partial t},\hspace{0.8cm}\Gamma^0_{01}=-\Gamma^1_{11}=\frac{1}{2\alpha}\frac{\partial\alpha}{\partial x},\hspace{0.8cm}\Gamma^0_{11}=-\frac{1}{2\alpha^3}\frac{\partial\alpha}{\partial t},\hspace{0.8cm}\Gamma^1_{00}=\frac{1}{2}\alpha\frac{\partial\alpha}{\partial x} ,
\edm
the only independent component of the Riemann tensor and the curvature scalar are
\bdm
2\alpha R^0_{\;101}=R=\frac{\partial}{\partial t^2}\left(\frac{1}{\alpha}\right)-\frac{\partial\alpha}{\partial x^2}
\edm
and the diagonal elements of the energy-momentum tensor of matter are
\beqr
T_{00}&=&\frac{1}{2}\left(\frac{\partial\phi}{\partial t}\right)^2+\frac{1}{2} \alpha^2\left(\frac{\partial\phi}{\partial x}\right)^2+\alpha V(\phi)\label{t00}\\
T_{11}&=&\frac{1}{2\alpha^2}\left(\frac{\partial\phi}{\partial t}\right)^2+\frac{1}{2} \left(\frac{\partial\phi}{\partial x}\right)^2-\frac{1}{\alpha} V(\phi).\label{t11}
\eeqr
Thus the field equations (\ref{eqjt1})-(\ref{eqjt2}) take the form
\beqr
-\frac{\partial}{\partial t}\left(\frac{1}{\alpha}\frac{\partial\phi}{\partial t}\right)+\frac{\partial}{\partial x}\left(\alpha\frac{\partial\phi}{\partial x}\right)&=&\frac{dV}{d\phi}\label{eqjt1gauge}\\
\frac{\partial^2}{\partial t^2}\left(\frac{1}{\alpha}\right)-\frac{\partial^2\alpha}{\partial x^2}-\Lambda&=&-16\pi G V(\phi).\label{eqjt2gauge}
\eeqr
Henceforth, we will concentrate on static configurations where both $\alpha$ and $\phi$ depend only on the space-like coordinate $x$. Also, for simplicity, we will be concerned only with the atractive JT gravitational force and will set the cosmological constant to zero. The reason is  that our main interest is to explore the behavior, once gravitation is put on, of some standard kink solutions that are usually studied on a 1+1 dimensional Minkowski setting, instead of its de Sitter or anti-de Sitter counterparts, which are by themselves non-static geometric backgrounds. Also, for concreteness, in the rest of this section we will focus on the the simplest and most paradigmatic model exhibiting spontaneous symmetry breaking, the $\phi^4$ theory with potential
\beq
V(\phi)=\frac{\lambda}{4}(\phi^2-v^2)^2.\label{potencialphi4}
\eeq
Let us mention that in the static case it is also useful another gauge for the metric, namely $ds^2=-\gamma(z) dt^2+dz^2$, a gauge that has been used, for instance, to study stellar structure in two dimensions in \cite{SiMa91}. Of course, we can recover this gauge from our present conventions by defining a spatial coordinate $z$ by $dz=\frac{dx}{\sqrt{\alpha(x)}}$ and then $\gamma(z)=\alpha(x(z))$.

The theory with potential (\ref{potencialphi4}) has been much investigated, both from the point of view of elementary particle physics, where renormalizability in four dimensions makes it especially interesting, and from a condensed matter perspective, where it is the typical interaction appearing in Ginzburg-Landau functionals. At the level of perturbation theory, the model encodes the interaction of real scalar quanta of mass $M=\sqrt{2\lambda} v$ through third and fourth order vertices, but the dynamics has also room for interesting non-perturbative phenomena. In fact, as it is well known, the space of finite-energy configurations splits in four sectors, which are classified by the topological charge $Q~=~\frac{1}{2 v}\int_{-\infty}^\infty \partial_x\phi$ and are both classically and quantum mechanically disconnected. There are two vacuum sectors in which the asymptotic values of the field are the same for positive and negative $x$, i.e., $\phi(t,\pm\infty)=v$ or  $\phi(t,\pm\infty)=-v$, along with other two non-trivial topological sectors with mixed asymptotics given by $\phi(t,\pm\infty)=\pm v$ or $\phi(t,\pm\infty)=\mp v$. The decay of a configuration with non-vanishing topological charge to one of the two constant classical vacua $\phi=-v$ or $\phi=v$ is forbidden by the presence of infinite potential barriers among the sectors. Thus, in the topological sectors the energy is minimized by a kink or antikink, a static solution of the Euler Lagrange equations which continuously interpolates between different vacua. Indeed, we see from (\ref{t00}) that in Minkowski spacetime the energy of a static configuration can be written in the form
\bdm
E[\phi]=\frac{1}{2}\int_{-\infty}^\infty\left[\frac{d\phi}{dx}\pm\sqrt{\frac{\lambda}{2}}(\phi^2-v^2)\right]^2\mp\sqrt{\frac{\lambda}{2}}\int_{\phi(-\infty)}^{\phi(\infty)} d\phi (\phi^2-v^2)
\edm
and this expression attains its minimum when the Bogomolnyi equation is satisfied
\bdm
\frac{d\phi}{dx}=\pm\sqrt{\frac{\lambda}{2}}(v^2-\phi^2).
\edm
The solutions corresponding to the topologically non-trivial asymptotic conditions are
\bdm
\phi(x)=\pm v\tanh\left(v\sqrt{\frac{\lambda}{2}}(x-x_0)\right),
\edm
where the plus and minus signs correspond, respectively, to the kink $\phi_K(x)$ and antikink $\phi_{AK}(x)$ configurations. Notice that the Bogomolny equation implies the Euler-Lagrange equation
\bdm
\frac{d^2\phi}{dx^2}=\mp \sqrt{2\lambda}\;\phi\;\frac{d\phi}{dx}=\lambda\phi(\phi^2-v^2)
\edm
and the kink and antikink are thus true solutions of the theory. Their energy is
\bdm
E[\phi_K]=E[\phi_{AK}]=\frac{2}{3}\sqrt{2\lambda}\;v^3.
\edm

For definiteness, to study the effect of Jackiw-Teitelboim gravity on configurations of this type we will focus on kinks rather than on antikinks, which are analogous. Thus, we have to solve the field equations (\ref{eqjt1gauge})-(\ref{eqjt2gauge}) in the static limit, with $\Lambda$ set to zero and boundary conditions $\phi(-\infty)=-v$, $\phi(+\infty)=v$. In two dimensions, the scalar field $\phi$ and the constants $v$ and $G$ are non-dimensional quantities, while the coupling $\lambda$ has mass dimension two. It is convenient to shift to a set of non-dimensional variables by redefining
\bdm
\phi=v\psi,\hspace{1cm}x=\sqrt{\frac{2}{\lambda}}\frac{y}{v},\hspace{1cm}8\pi G v^2=g,
\edm
so that the equations become 
\beqr
\frac{d}{d y}\left(\alpha\frac{d\psi}{dy}\right)&=&2\psi(\psi^2-1)\label{eqk1a}\\
\frac{d^2\alpha}{dy^2}&=&g(\psi^2-1)^2.\label{eqk1b}
\eeqr
By translational invariance we can take the center of the kink, i.e., the point at which $\psi$ is zero, at the origin of the $y$ coordinate. Of course, in this case, given the $\mathbb{Z}_2$ symmetry of the potential and the form of the kink boundary conditions, the kink profile has to be an odd function of $y$. Then the equations imply that the metric coefficient $\alpha$ is even and, in order to completely fix the set up, we can define the timelike variable $t$ in such a way that it measures the proper time at the core of the kink. It thus follows that it is enough to solve (\ref{eqk1a})-(\ref{eqk1b}) for $y\geq 0$ and with boundary conditions at $y=0$ given by
\beqr
&&\alpha(0)=1,\hspace{2cm}\left.\frac{d\alpha}{dy}\right|_{y=0}=0,\label{boundk1a}\\
&&\psi(0)=0,\hspace{2cm}\left.\frac{d\psi}{dy}\right|_{y=0}=p,\label{boundk1b}
\eeqr
where the slope $p$ of the kink profile $\psi$ at the origin has to be chosen in such a way that the boundary condition at infinity
\beq
\psi(+\infty)=1\label{asympk1}
\eeq 
is satisfied.
 
Once the equations are solved, the solution can be used to compute several physical quantities characterizing the kink. Among these, the most  relevant ones are the total rest energy, the energy density and the pressure distribution inside the kink. The total energy is 
\bdm
E[\phi]=\int_{-\infty}^{\infty} dx \sqrt{\alpha(x)}\; T^{00}(x)=v^3\sqrt{\frac{\lambda}{2}} E_{\rm norm}[\psi]
\edm
where the ``normalized" non-dimensional rest energy $E_{\rm norm}[\psi]$ is
\beq
E_{\rm norm}[\psi]=\int_0^\infty dy \left\{\sqrt{\alpha}\left(\frac{d\psi}{dy}\right)^2+\frac{1}{\sqrt{\alpha}}(\psi^2-1)^2\right\}.
\label{enerk1}
\eeq
Thus, the energy density of the kink, also in normalized form, is
\bdm
{\cal E}_{\rm norm}(y)=\sqrt{\alpha(y)}\left(\frac{d\psi}{dy}\right)^2+\frac{1}{\sqrt{\alpha(y)}}\left[\psi^2(y)-1\right]^2,
\edm
where, since we integrate only from zero to infinity, our normalization includes in this case a factor of two with respect to the true energy density. The pressure, on the other hand, is given by ${\cal P}=\frac{T^{11}}{\alpha}=\frac{\lambda v^4}{4} {\cal P}_{\rm norm}$, with the normalized pressure distribution being
\beq
{\cal P}_{\rm norm}(y)=\alpha(y) \left(\frac{d\psi}{dy}\right)^2-\left[\psi^2(y)-1\right]^2 \label{prek1}.
\eeq
\subsection{Some numerical results}
If we solve the system (\ref{eqk1a})-(\ref{asympk1}) with vanishing $g$, the gravitational field decouples, the background geometry is 1+1 dimensional Minkowski spacetime and we recover the standard kink of $\phi^4$ theory, which in rescaled variables reads
\bdm
\alpha(y)=1\hspace{2cm} \psi(y)=\psi_K(y)=\tanh(y).
\edm
In particular, the slope $p$ at the origin is unity. Also, we can compute the normalized energy, energy density and pressure to find the results
\bdm
E_{\rm norm}[\psi_K]=\frac{4}{3}\hspace{2cm} {\cal E}_{\rm norm}^{\psi_K}(y)= 2\;\sech^4 (y)\hspace{2cm}{\cal P}_{\rm norm}^{\psi_K} (y)=0.
\edm
Now, we turn the gravitational interaction on. Since the system (\ref{eqk1a})-(\ref{eqk1b}) cannot be analytically solved for arbitrary values $g>0$, we have to resort to approximate methods. Thus, after analyzing the behavior of the kink fields both inside its core and at large distances, we have to carry out a numerical integration of the field equations, seeking for a consistent interpolation between these regions. The boundary conditions (\ref{boundk1a})-(\ref{boundk1b}) imply that the metric coefficient near the origin has the form $\alpha(y)\simeq 1+\frac{1}{2} g y^2$ and substitution in (\ref{eqk1a}) gives, to dominant order, the differential equation
\bdm
\frac{d^2\psi}{dy^2}+g y \frac{d\psi}{dy} +2\psi=0\hspace{2cm}y\simeq 0.
\edm
With the change of variables $2 z=-g y^2$, this is a Kummer equation $z\frac{d^2\psi}{dz^2}+(\frac{1}{2}-z)\frac{d\psi}{dz}-\frac{1}{g}\psi=0$,  which, along with (\ref{boundk1b}), determines the form of $\psi$ near the center of the kink as
\bdm
\psi(y)= p y\,  {}_1 \! F_1(\frac{1}{2}+\frac{1}{g}; \frac{3}{2}; -\frac{g y ^2}{2})\simeq p y-\frac{1}{6} p(g+2) y^3+\ldots\hspace{2cm} y\simeq 0
\edm
where ${}_1 \! F_1(a;b;z)$ is the confluent hypergeometric function of the first kind. 

\begin{figure}
\centering
\includegraphics[width=8cm]{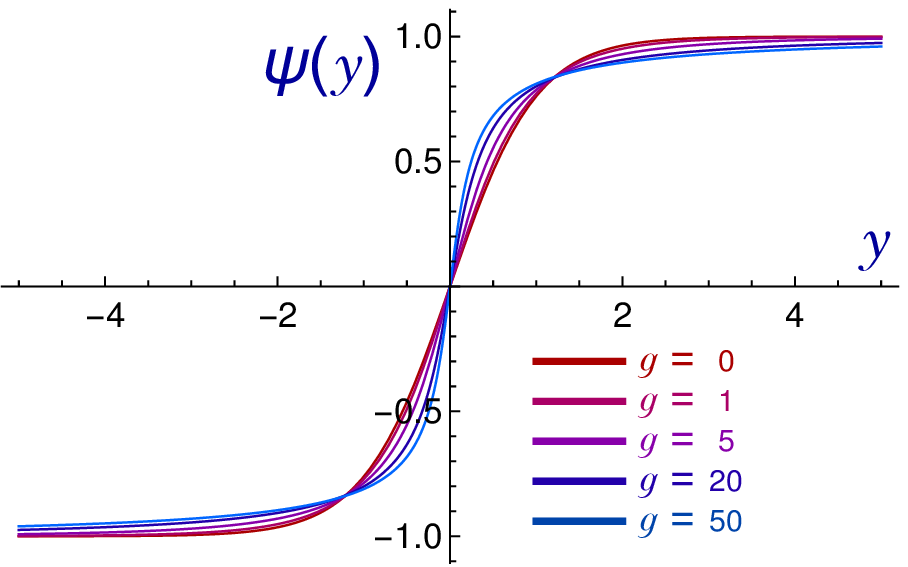}
\hspace{1cm}
\includegraphics[width=8cm]{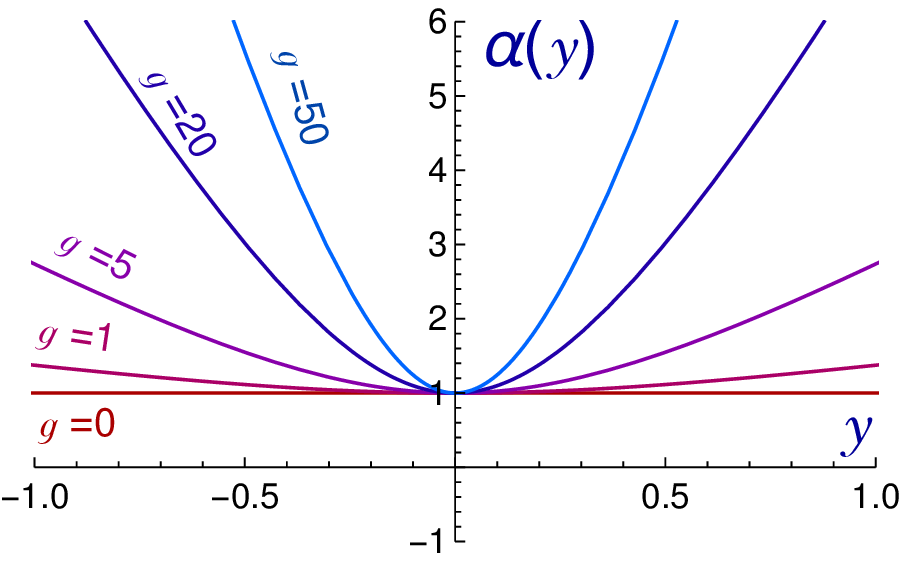}
\caption{$\psi$ and $\alpha$ profiles of the $\phi^4$ kink for several strengths of the gravitational interaction.}
\label{fig1}
\end{figure}

\begin{figure}
\centering
\includegraphics[width=8cm]{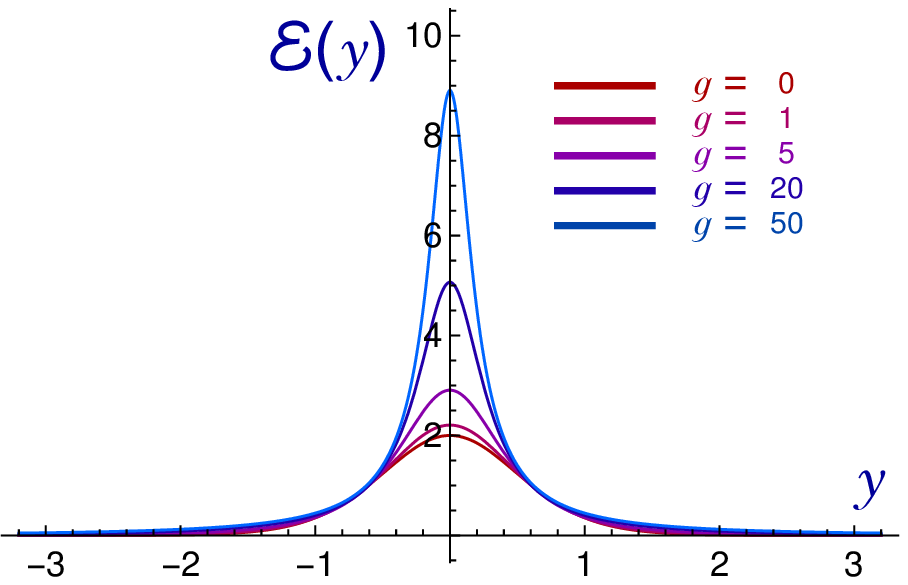}
\hspace{1cm}
\includegraphics[width=8cm]{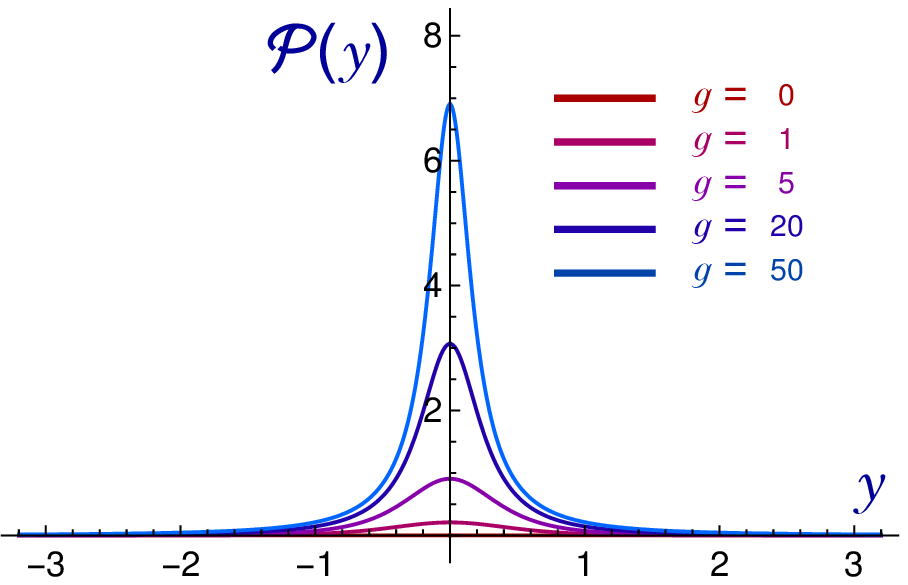}
\caption{The normalized energy density and pressure distribution of  the $\phi^4$ kink for several for several strengths of the gravitational interaction.}
\label{fig2}
\end{figure}

In the asymptotic region, on the other hand, we write $\psi(y)=1-\xi(y)$ and work at leading order in $\xi$. Equation(\ref{eqk1b}) and boundary condition (\ref{asympk1}) demand that
\bdm
\alpha(y)=q y+r\hspace{2cm}y\rightarrow\infty
\edm
for some coefficients $q$ and $r$. Using this expression in (\ref{eqk1a}) leads to the differential equation
\[
qy \frac{d^2 \xi}{dy^2}+q\frac{d\xi}{dt}-4\xi=0 \hspace{0.5cm} y\rightarrow \infty
\]
Therefore, the solution is
\[
\xi(y)=a K_0 \Big( \frac{4\sqrt{y}}{\sqrt{q}}  \Big) + b I_0 \Big( \frac{4\sqrt{y}}{\sqrt{q}}  \Big) \approx \frac{1}{2\sqrt{2}} \Big(  \frac{q}{y} \Big)^{\frac{1}{4}} \Big( a \sqrt{\pi} e^{-\frac{4\sqrt{y}}{\sqrt{q}}} +  \frac{b}{\sqrt{\pi}} e^{\frac{4\sqrt{y}}{\sqrt{q}}} \Big) \hspace{0.5cm} y\rightarrow \infty .
\]
Hence, there should be a solution to (\ref{eqk1a})-(\ref{eqk1b}) on the half-line $[0,+\infty)$ with this asymptotic behavior, and a critical value $p=p_{\rm crit}$ of the slope at the origin such that the coefficient $b$ vanishes, thus fulfilling the proper boundary condition (\ref{asympk1}). The task at hand is to integrate numerically the system (\ref{eqk1a})-(\ref{eqk1b}) while varying the value of $p$ until the critical value is attained. Hence, we use a shooting method, starting from $y=0$, integrating the equations for positive values of $y$ and then extending the solution to negative values by means of the known parities of $\psi(y)$ and $\alpha(y)$. We look for the value of the slope at the origin which gives a monotonically increasing $\psi(y)$ that asymptotically reaches $\psi=1$. For all values of the coupling, we obtain convergence to a convincing kink profile in which $\psi$ takes a constant value very close to one for an ample interval, with width of order $\Delta y\simeq 10$ for small $g$ up to $\Delta y\simeq 40$ for higher $g$, much larger than the size of the core of the kink. We thus expect that the $b$ coefficient multiplying the term which makes $\psi$ to diverge exponentially from the correct asymptotic value is very small and the solutions found are accurate. We have performed this for different values of the coupling $g$ and we show the results in Figure \ref{fig1}, where the profiles for $\psi(y)$ and $\alpha(y)$ are displayed, and in Figure \ref{fig2}, which shows the normalized energy density and pressure distributions of the kinks.
Also, we list in Table \ref{table1} the values of $p_{\rm crit}$; the normalized energy; the maxima of the energy density and pressure distributions, which, of couse, are located at the center of the kink; the width of the kink, which we have conventionally defined as the value of the $y$ coordinate containing a 95\% of the total energy; the slope $q$ and intercept $r$ of the asymptotic $\alpha$ profile and, finally, a parameter $\kappa$ whose meaning is explained below.
\begin{table}[t]
\begin{center}
\begin{tabular}{||c|c|c|c|c|c|c|c|c||}
\hline
\multicolumn{9}{||c||}{Numerical results for $\phi^4$ kinks}\\
\hline
$g$&$p_{\rm crit}$&$E_{\rm norm}$&${\cal E}_{\rm norm}(0)$&${\cal P}_{\rm norm}(0)$&width&$q$&$r$&$\kappa$\\
\hline
0.00&1.000&1.333&2.000&0.000&1.131&0.000&1.000&1.000\\
\hline
0.10&1.011&1.333&2.022&0.022&1.129&0.066&0.971&0.996\\
\hline
0.50&1.052&1.335&2.107&0.107&1.123&0.327&0.854&0.980\\
\hline
1.00&1.099&1.338&2.208&0.208&1.117&0.645&0.710&0.963\\
\hline
3.00&1.254&1.361&2.574&0.574&1.099&1.855&0.132&0.909\\
\hline
5.00&1.381&1.389&2.907&0.907&1.084&3.011&-0.472&0.867\\
\hline
10.00&1.634&1.460&3.670&1.670&1.050&5.780&-2.143&0.792\\
\hline
20.00&2.016&1.592&5.066&3.066&0.987&11.074&-6.233&0.696\\
\hline
30.00&2.320&1.709&6.382&4.382&0.932&16.214&-11.334&0.633\\
\hline
40.00&2.580&1.815&7.656&5.656&0.884&21.269&-17.425&0.586\\
\hline
50.00&2.811&1.912&8.903&6.903&0.842&26.267&-24.460&0.549\\
\hline
\end{tabular}
\end{center}
\caption{Some parameters characterizing the $\phi^4$ kink for different values of $g$}
\label{table1}
\end{table}

As one can see from the figures, the rate at which the scalar field varies behaves differently near the origin and afar from it. In the central region, the kink profile is steeper for higher $g$ but, as $y$ increases, the kinks of high gravitational coupling approach the minimum of the potential more slowly than those with small $g$. The energy density and the pressure decrease quickly to zero from their maxima at the core of the kink, and both the values of these maxima and the rate of decreasing are grater as the parameter $g$ increases. In fact, as it can be checked from the table, the total energy of the kink increases with $g$, while the trend for its size variation goes in the opposite sense: the more the coupling $g$ grows, the more the energy is concentrated around the origin, as it should be expected under the influence of a gravitational field. Indeed, as it is shown in the figures, the metric distortion created by the kink at its core is very pronounced, except for very small $g$. On the other hand, although as $y$ increases and $\alpha(y)$ approaches its linear regime the curvature tends to vanish, there is, however, a residual gravitational force at great distance. Mechanical energy conservation for a particle of rescaled non-dimensional mass $m$ falling from rest at $y=y_0$ under the gravity of the kink implies that velocity and acceleration are 
\bdm
v(y)=\alpha(y)\sqrt{1-\frac{\alpha(y)}{\alpha(y_0)}},\hspace{2cm}a(y)=\alpha(y)\left(1-\frac{3\alpha^2(y)}{2\alpha(y_0)}\right)\left.\frac{d\alpha}{dy}\right|_{y},
\edm
and thus the gravitational force on a particle at rest at position $y$ is
\bdm
F\equiv m a=-\frac{1}{2} m\alpha(y)\left.\frac{d\alpha}{dy}\right|_y\rightarrow -\frac{1}{2} m q y\hspace{1cm} {\rm for}\ y\rightarrow\infty,
\edm
where the values of $q$ can be read from the table.

Regarding the interpretation of last column in Table \ref{table1}, let us recall that for a point particle of mass $M$ located at the origin, with the stipulation $\alpha(0)=1$ that we are sticking to, the metric coefficient, in terms of the dimensional $x$ coordinate, is $\alpha(x)=4\pi G M|x|+1$ \cite{SiMa91}. Thus, transforming this formula to non-dimensional variables, we can write the relationship between the point-particle mass and the slope of the metric coefficient at infinity as
\bdm
\left.\frac{d\alpha}{dy}\right|_{y=\infty}=\frac{4\sqrt{2}\pi G M}{v\sqrt{\lambda}}.
\edm
For the kink, on the other hand, from (\ref{eqk1b}) and (\ref{boundk1a}), we have
\bdm
\left.\frac{d\alpha}{dy}\right|_{y=\infty}=8\pi G v^2 I[\psi]\hspace{2cm} I[\psi]=\int_0^\infty dy (\psi^2-1)^2.
\edm
Thus, from the point of view of the the long distance analysis, in which the kink appears as a point particle, it makes sense to assign to the kink a gravitational mass 
\bdm
M_{\rm GRAV}=2 v^3 \sqrt{\frac{\lambda}{2}} I[\psi].
\edm
Nevertheless, looking at (\ref{enerk1}) we see that the inertial mass is different:
\bdm
M_{\rm INER}=E[\psi]=v^3\sqrt{\frac{\lambda}{2}} E_{\rm norm}[\psi].
\edm
The last column in Table \ref{table1} gives account of this difference by means of the parameter $\kappa$, which is, precisely, the quotient between $M_{\rm GRAV}$ and $M_{\rm INER}$, i.e., $\kappa=2 \frac{I[\psi]}{E_{\rm norm}[\psi]}$. The discrepancy between both mass values stems from the fact that, in contrast with the case of a true point particle, for an extended object such as the kink the two diagonal elements $T_{00}$ and $T_{11}$ of the energy-momentum tensor, i.e. not only energy density but also pressure, source JT gravity, having thus an effect on the long distance metric. The situation is analogous of what happens in general relativiy in 3+1 dimensions, where the exterior metric of a ball of perfect fluid in equilibrium is the Schwarzschild solution with a mass parameter $m$ which differs from the mass $m^\prime$ obtained by integrating the energy density of the fluid in the ball, see for instance \cite{Carr04}, an effect which is interpreted as due to the existence of a binding energy originated for the attractive gravitational forces in the interior of the fluid, which are, in fact, compensated by the pressure. For the standard kink, $\frac{d\psi_K}{dy}=1-\psi_K^2$ and $\alpha(y)=1$, thus $E_{\rm norm}[\psi_K]=2 I[\psi_k]$ and $\kappa=1$, consistently with the fact that for $g\rightarrow 0$ internal gravitational self-energy and pressure disappear. For other values of $g$, the numerical computation of the integrals gives the results collected in the table.
\subsection{The limit of small gravitational coupling: perturbative analysis}
\begin{figure}[t]
\centering
\includegraphics[height=5cm]{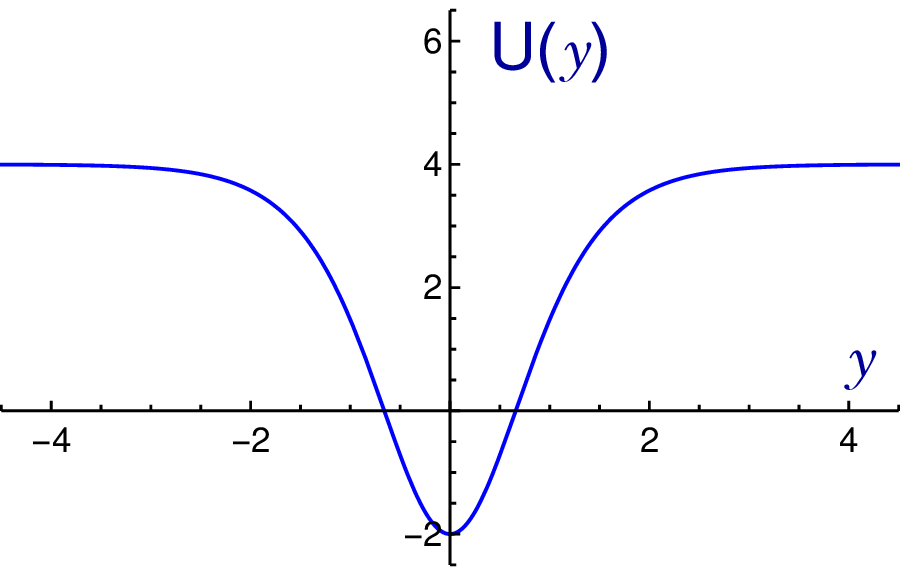}
\hspace{0.5cm}
\includegraphics[height=5cm]{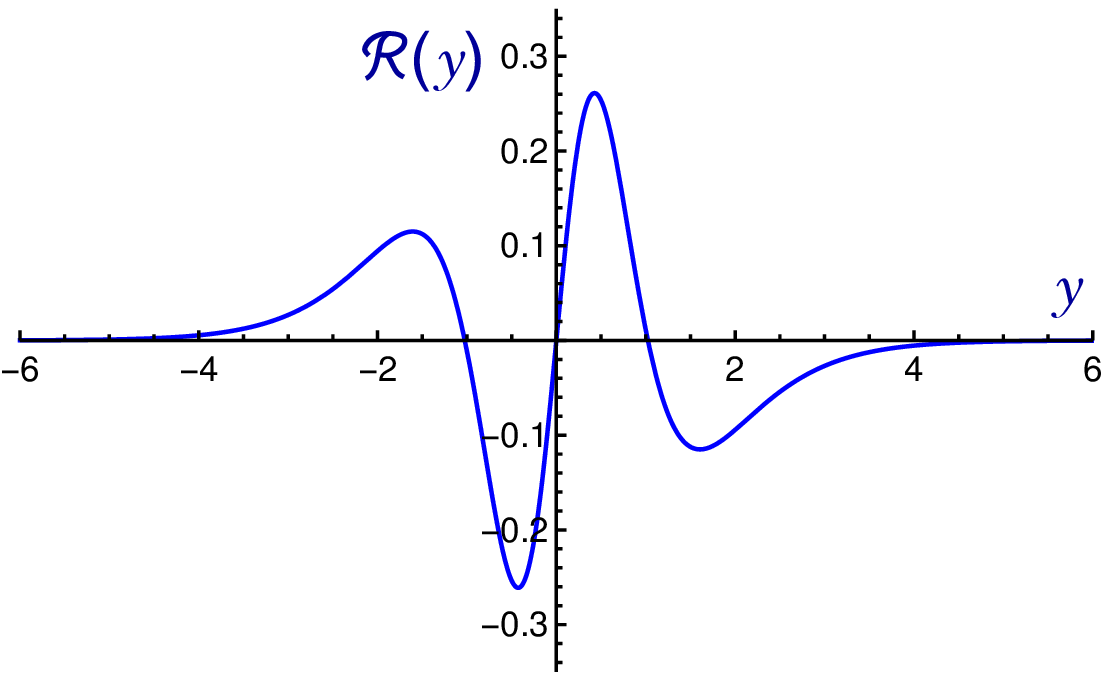}
\caption{The potential well $U(y)$ and the source ${\cal R}(y)$ of the inhomogeneous Schr\"{o}dinger equation for the $\phi^4$ kink.}
\label{fig3}
\end{figure}

We have demonstrated the action of JT gravity on the kinks of the of $\phi^4$ theory, allowing for the possibility  that the gravitational coupling can be, in principle, arbitrary high. Of course, this is in contrast with the usual point of view adopted when this theory, or other quantum field theories, are investigated which assumes that, compared with the other interactions present in the system, gravity is so weak that its effects can be completely disregarded. This approach has brought out a plenty of remarkable results on kinks or other topological defects. Thus, in the event that one is interested in reintroducing gravity in the picture, it is quite reasonable to limit the shifting of the gravitational coupling from zero to a tiny value.  In this subsection, we will adopt this perspective by taking $g$ small enough to make a linear perturbative analysis of the kink equations feasible. As we know, for $g=0$ the solution is the standard kink and the spacetime is Minkowski. Hence, we shall assume that
\beqr
\psi(y)&=&\psi_K(y)+g\varphi(y)+o(g^2)\label{expa}\\
\alpha(y)&=&1+g \beta(y)+o(g^2)\label{expb}
\eeqr
and we will substitute these expressions in (\ref{eqk1a})-(\ref{eqk1b}) keeping only the linear terms in the coupling. Given that $\alpha(y)$ increases monotonically for $y\rightarrow\infty$, perturbation theory breaks down at great distances, but we see from Figure 1 that if $g$ is small enough there is a wide interval around the core of the kink in which $\alpha(y)\simeq 1$ and, at least in this region, the perturbative treatment should give a good approximation of the complete solution. Having in mind this point and plugging the expansions (\ref{expa}) and (\ref{expb}) into  equations (\ref{eqk1a}) and (\ref{eqk1b}),  we obtain the following linear equations
\beqr
{\cal H}_y \varphi&=&\frac{d}{dy}\left(\beta\frac{d\psi_K}{dy}\right)\label{eqk1perta}\\
\frac{d^2\beta}{dy^2}&=&\left(\psi_K^2-1\right)^2\label{eqk1pertb}
\eeqr
where ${\cal H}_y$ is the Hessian operator in the background of the standard kink solution, whose form is
\bdm
{\cal H}_y=-\frac{d^2}{dy^2}+U(y)\hspace{2cm}U(y)=2\left(3\psi_K^2(y)-1\right).
\edm 
The boundary conditions at the origin (\ref{boundk1a})-(\ref{boundk1b}), on the other hand, become in this regime
\beqr
\beta(0)=0\hspace{2cm}\left.\frac{d\beta}{dy}\right|_{y=0}=0\label{boundk1perta}\\
\varphi(0)=0\hspace{2cm}\left.\frac{d\varphi}{dy}\right|_{y=0}=s\label{boundk1pertb}
\eeqr
with the functions $\beta(y)$ and $\varphi(y)$ being, respectively, even and odd in $y$. The value of $s$ has to be chosen in such a way that the asymptotic behavior of $\varphi(y)$ is consistent with (\ref{asympk1}), which implies
\beq
\varphi(\infty)=0\label{asympk1pert}.
\eeq
Equation (\ref{eqk1pertb}) combined with the boundary condition (\ref{boundk1perta}) allows for a direct computation of the perturbation of the metric coefficient
\beq
\beta(y)=\int_0^y dz\int_0^z du\; {\rm sech}^4(u)=\frac{1}{6}\left\{1+4\log\left(\cosh(y)\right)-\sech^2(y)\right\}.
\eeq
As it should be, this is a function interpolating between a parabola at the core of the kink and a straight line at large distances, in fact
\beqrn
\beta(y)&\simeq&\frac{y^2}{2}\hspace{5cm}y\simeq 0\\
\beta(y)&\simeq& \frac{2}{3} y+\frac{1-4\log 2}{6}\hspace{2.3cm}y\rightarrow \infty
\eeqrn
Consequently, this computation provides us with exact values, in the limit of small $g$, for the $q$ and $r$ coefficients that we had computed numerically in the previous subsection. The other equation, (\ref{eqk1perta}), is a Schr\"{o}dinger equation of non-homogeneous type. The potential $U(y)$ of the Schr\"{o}dinger operator is a symmetric well
\bdm
U(y)=4-6\;\sech^2(y)
\edm 
which displays a minimum at the origin $U(0)=-2$ and attains a flat profile $U(y)\rightarrow 4$ when $|y|\rightarrow\infty$, see Figure \ref{fig3}. The source term, on the other hand, is
\beq
{\cal R}(y)=\frac{d}{dy}\left(\beta\frac{d\psi_K}{dy}\right)=\frac{1}{3}\left\{1-4\log\left(\cosh(y)\right)+2 \sech^2(y)\right\}\sech^2(y)\tanh(y),\label{rk1}
\eeq
an odd function whose behavior near the origin and at infinity is of the form
\beqr
{\cal R}(y)&\simeq&y\hspace{3.8cm}y\simeq 0\\
{\cal R}(y)&\simeq&-\frac{16}{3} y e^{-2|y|}\hspace{2cm}|y|\rightarrow \infty\label{rinfty}
\eeqr
and whose full profile is also shown in Figure \ref{fig3}.
\begin{figure}[b]
\centering
\includegraphics[height=5cm]{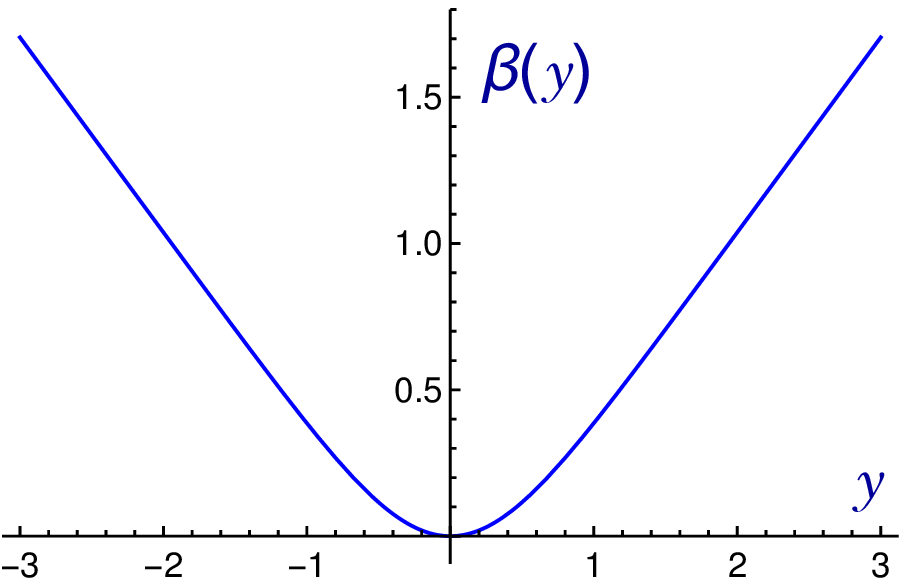}
\hspace{1cm}
\includegraphics[height=5cm]{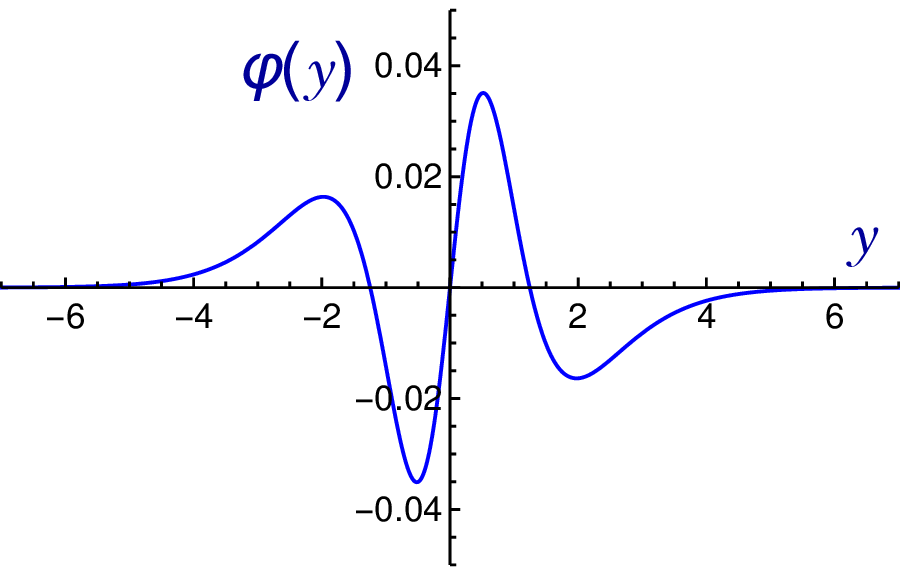}
\caption{The metric perturbation $\beta(y)$ and the response $\varphi(y)$ for the $\phi^4$ kink, this last one found by numerical methods.}
\label{fig3b}
\end{figure}
A numerical integration of the non-homogeneous Schr\"{o}dinger equation gives a graphic of $\varphi(y)$ as shown in Figure \ref{fig3b}, and it turns out that the value of $s$ leading to a good convergence at infinity is is $s\simeq 0.111$. However, the perturbative approach allows us to go beyond the numerical methods and to solve  exactly (\ref{eqk1perta}) in addition to (\ref{eqk1pertb}). Indeed, given the odd parity of $\varphi(y)$, it is enough to work out the solution for $y>0$. Thus, we can write
\beq
\varphi(y)=\int_0^\infty dz G(y,z) {\cal R}(z)\label{greenphi}
\eeq 
where $G(y,z)$ is a Green function in $[0,+\infty)$ of the Schr\"{o}dinger operator, 
\beq
\left[-\frac{d^2}{d y^2}+U(y)\right]G(y,z)=\delta(y-z)\label{eqgreenk1},
\eeq
with suitable boundary conditions to ensure that $\varphi(y)$ behaves at the origin and infinity as required by (\ref{boundk1pertb}) and (\ref{asympk1pert}). Notice that, although the Hessian operator of the standard non-gravitational kink has a normalizable zero mode due to translational invariance, this mode, being proportional to $\frac{d\psi_K}{dy}$, has no nodes. Thus, it satisfies the boundary condition at infinity, but not (\ref{boundk1pertb}). This fact guarantees that the Schr\"{o}dinger operator is invertible within the space of functions we are interested in, and therefore that the Green function we are seeking for exists. On the other hand, it is precisely by means of zero modes that this Green functions can be constructed, see for instance \cite{Col77}. In fact, it is not difficult to check that the general solution of (\ref{eqgreenk1}) is
\beq
G(y,z)=a(z)\rho(y)+b(z) \chi(y)+\frac{1}{W} \theta(y-z) \left[\chi(z)\rho(y)-\rho(z)\chi(y)\right]\label{gengreen}
\eeq
where $\rho(y)$ and $\chi(y)$ are, respectively, the even, normalizable, and odd, non-normalizable, zero modes of the Schr\"{o}dinger operator, $a(z)$ and $b(z)$ are arbitrary functions, $\theta(z)$ is the Heaviside step function and $W$ is the Wronskian of the zero modes:
\bdm
W=\rho \frac{d\chi}{dy}-\frac{d\rho}{dy}\chi.
\edm
The existence of two eigenmodes with zero eigenvalue and with the said features of normalizability and parity can be easily verified by taking advantage of the factorization of the Hessian operator,
\bdm
{\cal H}_y=D^\dagger_y D_y\hspace{2cm} D_y=\frac{d}{dy}+2 \tanh(y),
\edm
which implies that the normalizable mode comes about as the solution of
\bdm
D_y\rho(y)=0\Rightarrow \rho(y)=C_\rho\exp\left[-2\int_0^y du \tanh(u)\right],
\edm 
while the non-normalizable one follows from
\bdm
D_y^\dagger (D_y\chi)=0\Rightarrow D_y\chi=C_\chi\exp\left[2\int_0^y du \tanh(u)\right]
\edm
with $C_\rho$ and $C\chi$ arbitrary constants. Thus, the results are
\beqr
\rho(y)&=&\sech^2(y)\label{mod1k1}\\
\chi(y)&=&3y\;\sech^2(y)+\left(3+2 \cosh^2(y)\right) \tanh(y)\label{mod2k1},
\eeqr
where we have picked out a normalization such that $W=8$. Now, since $\rho(0)\neq 0$ and $\chi(0)=0$, we choose $a(z)=0$ in (\ref{gengreen}) in order that $G(0,z)=0$ for positive $z$, thus ensuring that  $\varphi(0)=0$ in accordance with (\ref{boundk1pertb}). As one can see, the remaining condition, (\ref{asympk1pert}), is satisfied by choosing $b(z)=\frac{1}{W}\rho(z)$. This leaves us with the integral formula
\beq
\varphi(y)=\frac{1}{W}\rho(y)\int_0^y dz \chi(z) {\cal R}(z)+\frac{1}{W}\chi(y)\int_y^\infty dz \rho(z) {\cal R}(z).\label{phi}
\eeq
In this expression, although $\chi(y)$ is non normalizable, the second term vanishes by construction when $y$ approaches infinity, whereas by virtue of the asymptotic behavior of the two zero modes of the kink
\bdm
\rho(y)\simeq 4 e^{-2y},\hspace{1cm}\chi(y)\simeq\frac{1}{2} e^{2y}\hspace{1cm}{\rm for}\hspace{1cm} y\rightarrow \infty,
\edm
and of ${\cal R}(y)$, written in (\ref{rinfty}) above, the first one goes as $y^2 e^{-2y}$ in the same limit. Thus $\varphi(y)$ vanishes for $y\rightarrow\infty$, as it should do. It turns out that, after plugging (\ref{rk1}) and (\ref{mod1k1})-(\ref{mod2k1}) into (\ref{phi}), the integrals can be done exactly, and the result is
\bdm
\varphi(y)=\frac{1}{6}\left(\frac{2}{3} \tanh(y)-{\rm Li}_2(-e^{-2y})-y(y-2\log 2)-\frac{\pi^2}{12}\right)\sech^2(y).
\edm
Here ${\rm Li}_n(y)$ is the polylogarithm function, which can be written as ${\rm Li}_n(y)=y \Phi(y,n,1)$ where $\Phi(y,n,v)$ is the Lerch transcendent function, see \cite{GrRy80} for more details about these functions. It is possible to check that the solution given coincides with the function drawn in Figure 4 and, in particular, the derivative at $y=0$ gives exactly $s=\frac{1}{9}$ as we had found numerically. Apart from that, other traits of the solution are that, as we can see in the figure, for $y\rightarrow +\infty$ the function reaches zero from below after an oscillatory regime near the origin, with a maximum value $\varphi=0.0351$ at $y=0.5176$ and a subsequent minimum at $y=1.976$ at which $\varphi=-0.0164$. The zero between these two extrema lies at $y=1.244$.

We have thus developed a successful perturbative approach for $g\rightarrow 0$. Let us now briefly comment on the opposite regime $g\rightarrow \infty$. In principle, one can proceed in the same way, defining $g=\frac{1}{g^\prime}$, obtaining the solution $\tilde{\psi}_K(y), \tilde{\alpha}_K(y)$ valid for $g^\prime=0$, and then perturbing this solution for small $g^\prime$, i.e. large $g$, as
\bdm
\psi(y)=\tilde{\psi}_K(y)+g^\prime \varphi(y)+\ldots,\hspace{1.5cm}\alpha(y)=\tilde{\alpha}_K(y)+g^\prime \beta(y)+\ldots .
\edm
Now, in order that the left hand side of equation (\ref{eqk1b}) remains finite, the kink solution for $g^\prime=0$ has to be singular, $\tilde{\psi}_K(y)={\rm sgn}(y)$, and, given that the kink core is now concentrated at $y=0$, the metric should be that of a point particle, $\tilde{\alpha}_K(y)=1+ D |y|$, where $D$ is a constant which reflects the gravitational effect of the singular kink. Substituting this in (\ref{eqk1a})-(\ref{eqk1b}) and working at leading order in $g^\prime$, we obtain a system of equations similar to (\ref{eqk1perta})-(\ref{eqk1pertb})
\beqrn
-\frac{d}{dy}\left(\tilde{\alpha}_K\frac{d\varphi}{dy}\right)+3\left(2\tilde{\psi}_K^2-1\right)\varphi&=&\frac{d}{dy}\left(\beta \frac{d \tilde{\psi}_K}{dy}\right)\\
\frac{d^2\beta}{dy^2}&=&4\tilde{\psi}_K^2\varphi^2,
\eeqrn
but now involving Dirac delta singularities due to the non-smooth character of both the unperturbed kink and the background metric. Also, from the figure \ref{fig1}, one should expect that the $D$ constant is divergent, and some regularization procedure, possibly involving a certain degree of arbitrariness, should accompany the present approach. We see thus that the case $g\rightarrow\infty$ is not so neatly defined as the situation for $g\rightarrow 0$ and its detailed investigation is the subject of a different work.
\section{Self-gravitating sine-Gordon solitons}
Besides the $\phi^4$ kink, the other most prominent example of a localized solution in a scalar relativistic field theory in 1+1 dimensions is the sine-Gordon soliton, which deserves this name, instead of the simple denomination of kink, due of its absolute stability after collisions; the $\phi^4$ kink is not a soliton in this sense, but only a solitary wave \cite{Ra82}. Like the $\phi^4$ theory, the sine-Gordon model has been intensely studied and applications for it have been found in a variety of fields ranging from the mechanics of coupled torsion pendula to the study of dislocations in crystals, Josephson junctions, DNA molecules or black holes, see for instance \cite{CuKeWi14,ZdSaDa13}. In what follows, we shall proceed along the lines developed in the previous section to transform the standard flat space sine-Gordon soliton into a self-gravitating object by coupling the field theory to JT gravity. 

The potential of the sine-Gordon Lagrangian is
\bdm
V(\phi)=\frac{\lambda}{\gamma^2}\left(\cos(\gamma\phi)+1\right)
\edm
and an important difference in the space of classical vacua with respect to the $\phi^4$ theory arises: now the degenerate vacua form an infinite set, $V(\phi)=0$ for $\phi=(2 n+1) \frac{\pi}{\gamma}$, $n\in\mathbb{Z}$, and the discrete symmetry is thus $\mathbb{Z}_n$ instead of $\mathbb{Z}_2$. Notwithstanding this, since finite energy static solutions of a real scalar field theory can interpolate only between consecutive vacua, for our purposes we can limit ourselves to consider configurations which approach asymptotically one of the two vacua $\phi=\pm \frac{\pi}{\gamma}$. In this way, we come back  to a situation analogous to that of $\phi^4$ theory. When gravity is neglected, we have a Bogomolny splitting, with Bogomolny equation
\bdm
\frac{d\phi}{dx}=\pm\sqrt{\frac{2\lambda}{\gamma^2}\left(\cos(\gamma\phi)+1\right)},
\edm
and kink-like and antikink-like solutions interpolating between that two vacua appear. The kink is the sine-Gordon soliton, with profile
\bdm
\phi_S(x)=\frac{4}{\gamma} \arctan\left(\tanh\left(\frac{\sqrt{\lambda} x}{2}\right)\right)Vachaspati2006
\edm
such that $\phi_S(\pm\infty)=\pm\frac{\pi}{\gamma}$. To study its generalization in the presence of the JT gravitational field, we rescale to non dimensional variables
\bdm
\phi=\frac{\psi}{\gamma},\hspace{1cm}x=\frac{y}{\sqrt{\lambda}},\hspace{1cm}16\pi \frac{G}{\gamma^2}=g
\edm
and, proceeding as we did in the $\phi^4$ theory, obtain field equations of the form
\beqr
\frac{d}{d y}\left(\alpha\frac{d\psi}{dy}\right)&=&-\sin\psi\label{eqk2a}\\
\frac{d^2\alpha}{dx^2}&=&g(\cos\psi+1).\label{eqk2b}
\eeqr
Due to the well definite parities in the variable $y$ of both $\psi$ and $\alpha$, the equations can be solved in $[0,+\infty)$ with the same boundary conditions (\ref{boundk1a}) and (\ref{boundk1b}) used before, whilst the asymptotic condition at infinity changes now to
\beq
\psi(+\infty)=\pi\label{asympk2}.
\eeq
The energy and pressure of the static solutions are $E[\phi]=\frac{\sqrt{\lambda}}{\gamma^2} E_{\rm norm}[\psi]$ and ${\cal P}=\frac{\lambda}{2\gamma^2}{\cal P}_{\rm norm}$, where the normalized quantities are those written in (\ref{enerk1}) and (\ref{prek1}) with $(\psi^2-1)^2$ replaced by $2(\cos\psi+1)$. In particular, the standard flat-space sine-Gordon soliton is
\bdm
\psi_S(y)=4\arctan(e^y)-\pi
\edm
and we find
\bdm
E_{\rm norm}[\psi_S]=8\hspace{1.7cm} {\cal E}_{\rm norm}^{\psi_S}(y)=8\,\sech^2(y)\hspace{1.7cm}{\cal P}_{\rm norm}^{\psi_S}(y)=0.
\edm

\begin{figure}[t]
\centering
\includegraphics[width=8cm]{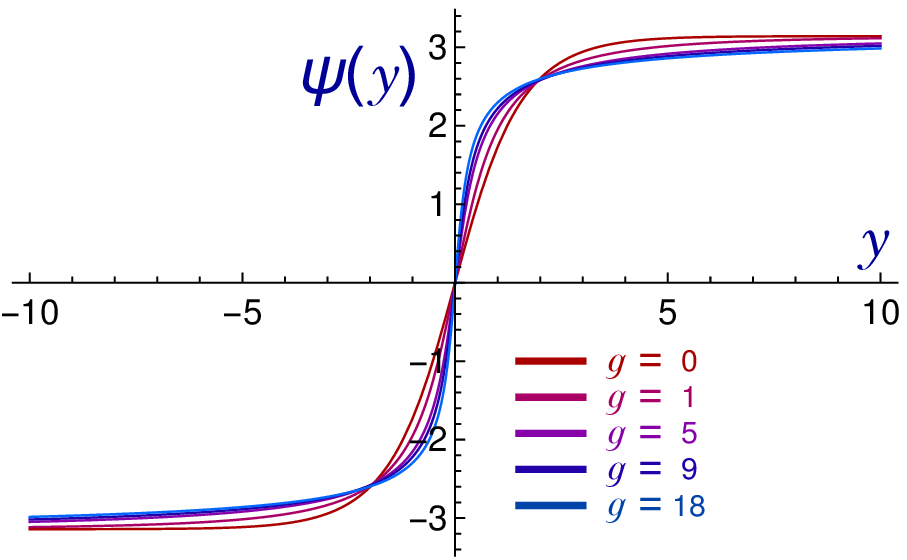}
\hspace{1cm}
\includegraphics[width=8cm]{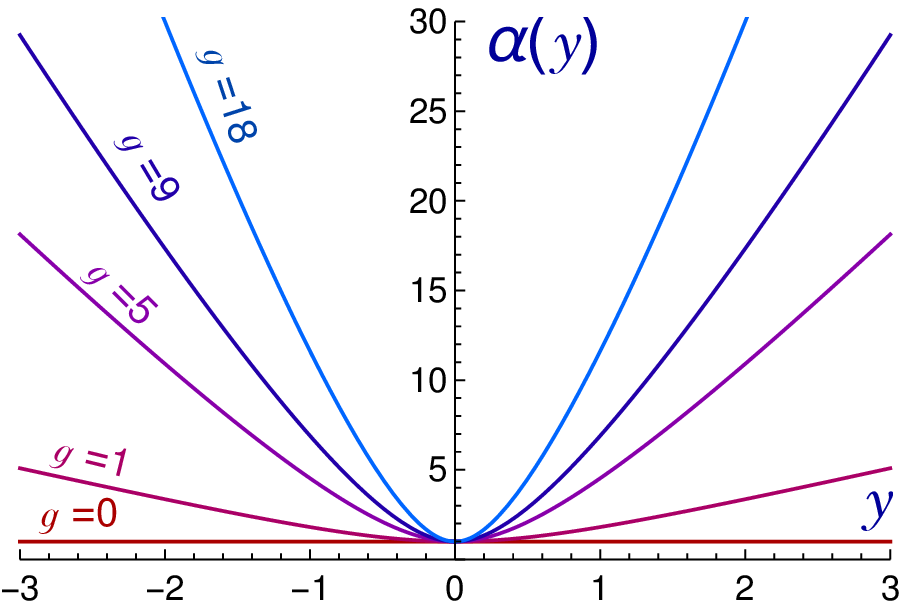}
\caption{$\psi$ and $\alpha$ profiles of the sine-Gordon soliton for several for several strengths of the gravitational interaction.}
\label{fig4}
\end{figure}

\begin{figure}[b]
\centering
\includegraphics[width=8cm]{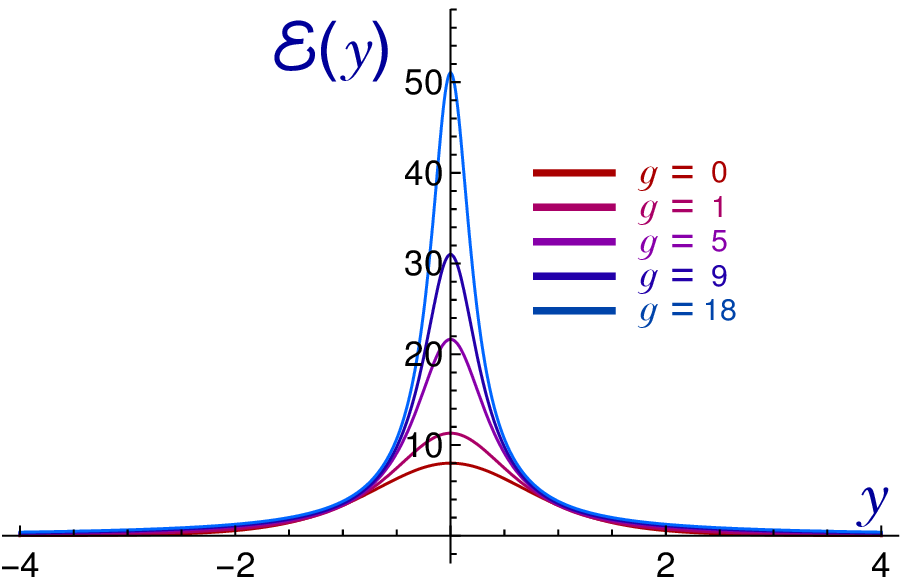}
\hspace{1cm}
\includegraphics[width=8cm]{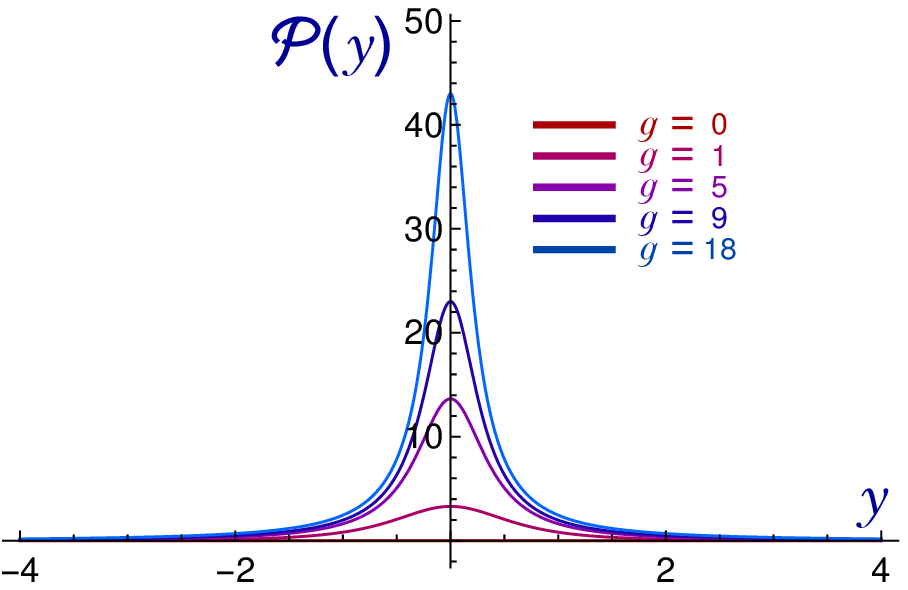}
\caption{The normalized energy density and pressure distribution of the sine-Gordon soliton for several for several strengths of the gravitational interaction.}
\label{fig5}
\end{figure}

As in Section 2, in order to look for numerical solutions of (\ref{eqk2a})-(\ref{eqk2b}) with the conditions (\ref{boundk1a}), (\ref{boundk1b}), (\ref{asympk2}), we first solve the system near the origin, to obtain
\beqrn
\alpha(y)&=&1+ g y^2\\
\psi(y)&=&p y\,  {}_1 \! F_1(\frac{1}{2}+\frac{1}{4g}; \frac{3}{2}; -g y ^2)\simeq p y-\frac{1}{6} p(2g+1) y^3+\ldots\hspace{2cm} y\simeq 0
\eeqrn
and also near infinity, where $\psi(y)=\pi-\xi(y)$ and the $\alpha$ profile is linear, $\alpha(y)=qy+r$. The behavior of $\xi$ turns out to be
\bdm
\xi(y)=a K_0 \Big( \frac{2\sqrt{y}}{\sqrt{q}}  \Big) + b I_0 \Big( \frac{2\sqrt{y}}{\sqrt{q}}  \Big) \approx \frac{1}{2} \Big(  \frac{q}{y} \Big)^{\frac{1}{4}} \Big( a \sqrt{\pi} e^{-\frac{2\sqrt{y}}{\sqrt{q}}} +  \frac{b}{\sqrt{\pi}} e^{\frac{2\sqrt{y}}{\sqrt{q}}} \Big) \hspace{0.5cm} y\rightarrow \infty 
\edm
and we have to  integrate the ODEs numerically to figure out the critical value of $p$ yielding $b=0$. Through this approach, we have found the results shown in the figures \ref{fig4} and \ref{fig5} and in Table \ref{table2}. The parameter $\kappa$ appearing in the table relates the gravitational and inertial masses of the soliton as in the previous section, and is now given by
\bdm
\kappa=4\frac{I[\psi]}{E_{\rm norm}[\psi]}\hspace{2cm}I[\psi]=\int_0^\infty dy [\cos(\psi)+1].
\edm\\\\
As the figures and table reflect, the qualitative features of the sine-Gordon solitons are similar to those already commented for the $\phi^4$ kinks. The most notorious difference is the that the values of energy, pressure and gravitational distortion are, with the natural normalization and non-dimensional variables that we are using, much greater for the former than for the latter.
  
\begin{figure}[b]
\centering
\includegraphics[height=5cm]{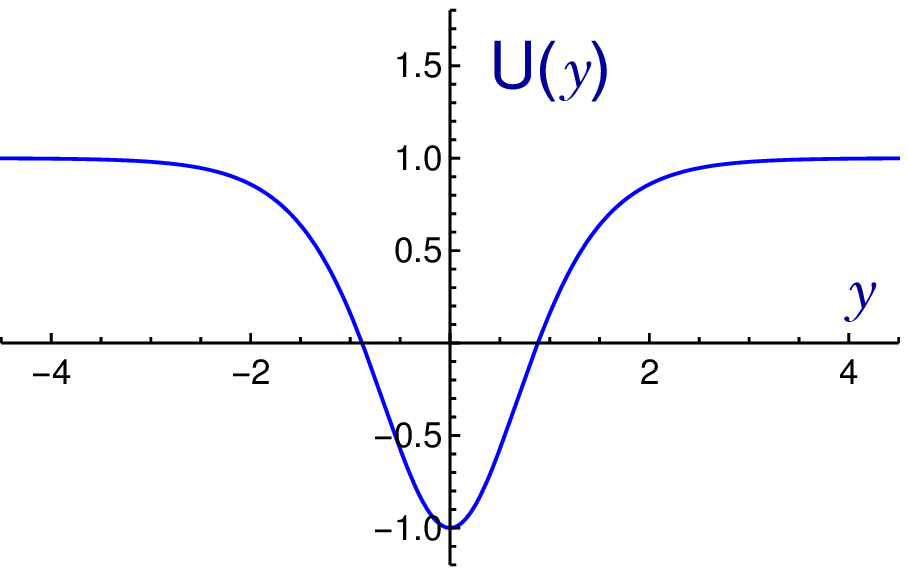}
\hspace{0.5cm}
\includegraphics[height=5cm]{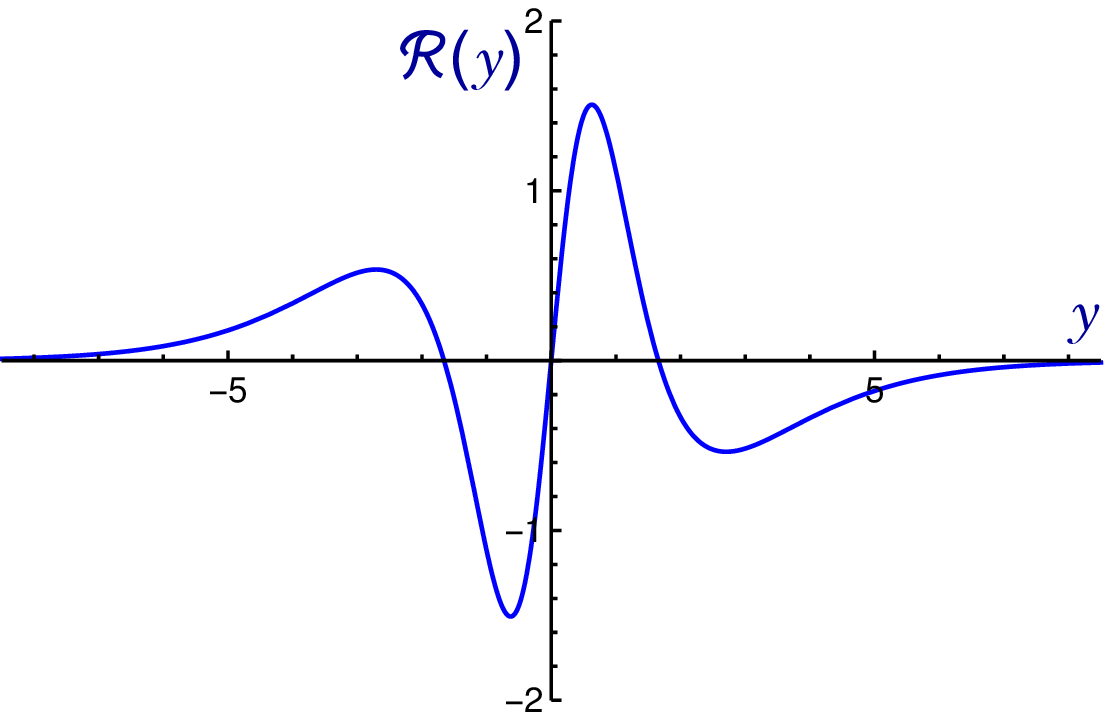}
\caption{The potential well $U(y)$ and the source ${\cal R}(y)$ of the inhomogeneous Schr\"{o}dinger equation for the sine-Gordon soliton.}
\label{fig6}
\end{figure}
\begin{table}[t]
\begin{center}
\begin{tabular}{||c|c|c|c|c|c|c|c|c||}
\hline
\multicolumn{9}{||c||}{Numerical results for sine-Gordon solitons}\\
\hline
$g$&$p_{\rm crit}$&$E_{\rm norm}$&${\cal E}_{\rm norm}(0)$&${\cal P}_{\rm norm}(0)$&width&$q$&$r$&$\kappa$\\
\hline
0.00 &2.000 &8.000 &8.000 &0.000 &1.832&0.000&1.000&1.000\\
\hline
0.10 &2.094 &8.008 &8.386 &0.386 &1.823&0.197&0.862&0.982\\
\hline
0.50 &2.401 &8.119 &9.766 &1.766 &1.800&0.940&0.307&0.926\\
\hline
1.00 &2.702 &8.309 &11.300 &3.302 &1.777&1.817&-0.437&0.875\\
\hline
3.00 &3.565 &9.097 &16.710 &8.707 &1.684&5.109&-4.155&0.749\\
\hline
5.00 &4.201 &9.808 &21.650 &13.650 &1.594&8.252&-9.075&0.673\\
\hline
7.00 &4.731 &10.450 &26.380 &18.380 &1.514&11.341&-15.507&0.620\\
\hline
9.00 &5.196 &11.040 &31.000 &23.000 &1.442&14.377&-22.929&0.579\\
\hline
11.00 &5.615 &11.580 &35.530 &27.530 &1.378&17.395&-31.867&0.546\\
\hline
15.00 &6.358 &12.580 &44.420 &36.420 &1.269&23.378&-53.842&0.496\\
\hline
18.00 &6.856 &13.260 &51.000 &43.000 &1.201&27.789&-71.581&0.466\\
\hline
\end{tabular}
\end{center}
\caption{Some parameters characterizing the solitons for different values of $g$.}
\label{table2}
\end{table}
The perturbative approach for the case of small $g$ can also be worked out like we did the case of $\phi^4$ theory. At first order in $g$, the equations for the perturbations $\beta$ and $\varphi$ of the flat metric and standard soliton are
\beqr
{\cal H}_y \varphi&=&\frac{d}{dy}\left(\beta\frac{d\psi_S}{dy}\right)\label{eqk2perta}\\
\frac{d^2\beta}{dy^2}&=&\cos(\psi_S)+1,\label{eqk2pertb}
\eeqr
where the Hessian operator is now
\bdm
{\cal H}_y=-\frac{d^2}{dy^2}+U(y)\hspace{2cm}U(y)=-\cos\left[\psi_S(y)\right]=1-2\, \sech^2(y),
\edm
whereas the boundary conditions are again (\ref{boundk1perta})-(\ref{asympk1pert}). Thus, integration of (\ref{eqk2pertb}) gives directly the answer for $\beta$
\bdm
\beta(y)=2 \log\left(\cosh(y)\right),
\edm
a function which interpolates consistently between a parabola and a straight line
\beqrn
\beta(y)&\simeq&y^2\hspace{4.2cm}y\simeq 0\\
\beta(y)&\simeq& 2 y-2\log 2\hspace{2.3cm}y\rightarrow \infty.
\eeqrn
The Hessian is a Schr\"{o}dinger operator with the potential drawn in Figure \ref{fig6}, a symmetric well showing a dip $U(0)=-1$ in the center and reaching asymptotic values $U(y)\rightarrow 1$ for $|y|\rightarrow \infty$. This operator is sourced by the function
\bdm
{\cal R}(y)=\frac{d}{dy}\left(\beta\frac{d\psi_S}{dy}\right)=4\left[1-\log\left(\cosh(y)\right)\right]\sech(y)\tanh(y),
\edm
which displays a linear behavior near the soliton center and decays exponentially at infinity
\beqrn
{\cal R}(y)&\simeq& 4y\hspace{3.3cm} y\simeq 0\\
{\cal R}(y)&\simeq&-8 y e^{-|y|}\hspace{2cm}|y|\rightarrow \infty,
\eeqrn
as it is also shown in the figure.
\begin{figure}[t]
\centering
\includegraphics[height=5cm]{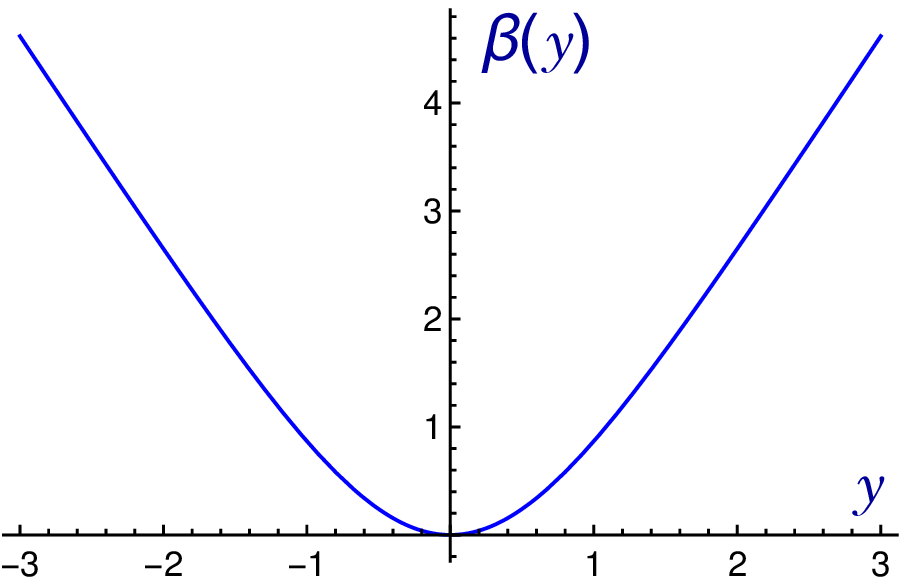}
\hspace{1cm}
\includegraphics[height=5cm]{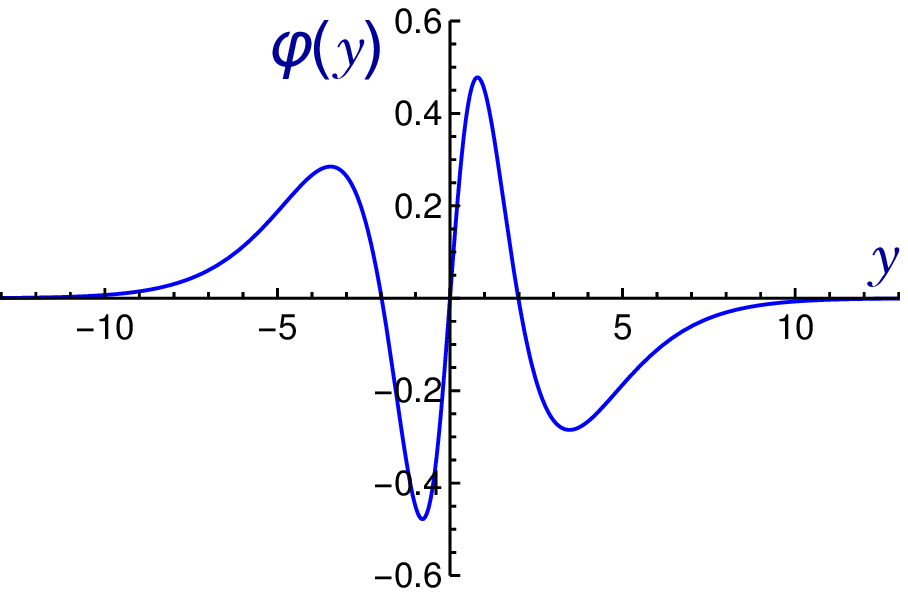}
\caption{The metric perturbation $\beta(y)$ and the response $\varphi(y)$ for the sine-Gordon soliton, this last one found by numerical methods.}
\label{fig6b}
\end{figure}
To compute $\varphi(y)$ we need to know the kernel of the Hessian. Again, this is facilitated by the factorization
\bdm
{\cal H}_y=D^\dagger_y D_y\hspace{2cm} D_y=\frac{d}{dy}+\tanh(y),
\edm
which allows for the computation the normalizable and non normalizable zero modes. They turn out to be
\beqrn
\rho(y)&=&\sech(y)\\
\chi(y)&=&y\;\sech(y)+\sinh(y),
\eeqrn
with Wronskian $W=2$. Substituting all that in (\ref{phi}) yields the final answer for the scalar perturbation of the soliton
\bdm
\varphi(y)=-\left({\rm Li}_2(-e^{-2y})+y(y-1-2\log 2)+\frac{\pi^2}{12}\right)\sech(y).
\edm
The shape of this function is portrayed in Figure \ref{fig6b}. In particular, taking the derivative at $y=0$ we learn that the parameter $s$ in the boundary condition (\ref{boundk1pertb}) is exactly one in the sine-Gordon theory. We can also obtain the upper and lower limits of the oscillation of the function as $y$ varies. There is a maximum at $y=0.797$ in which $\varphi=0.478$, and a minimum at $y=3.463$ in which $\varphi=-0.285$, with a zero in between at $y=1.981$. Thus, as it happened with the numerical results, in non dimensional variables the scale of the perturbative solution is considerably greater for the sine-Gordon case than for the $\phi^4$ one. 
\section{Self-gravitating kinks related to transparent P\"{o}schl-Teller potentials}
\subsection{Kink Hessian and supersymmetry}
In the two previous sections we have studied the coupling to JT gravity of the kink solutions of the $\phi^4$ and sine-Gordon models,  both for arbitrarily high gravitational coupling and in the regime of small $g$. Now, we will concentrate on the perturbative dominium only and will extend the treatment to deal with the self-gravitating kink solutions arising within an infinite hierarchy of field theoretical models, the P\"{o}schl-Teller hierarchy, which encompasses as particular cases the two theories considered so far. An important subject which shapes the hierarchy is unbroken supersymmetry.

As we have seen, a common feature of the $\phi^4$ and sine-Gordon models is that the Hessian operator of the standard, non gravitational, kink admit a factorization 
\beq
{\cal H}_y=D^\dagger_y D_y\hspace{2cm} D_y=\frac{d}{dy}+W(y)\label{factor}
\eeq
which, indeed, is the key instrument to obtain the perturbation of the scalar field once gravity is turned on. As it is well known, this factorization has a supersymmetrical origin. In fact, Witten $N=2$ supersymmetric quantum mechanics, see \cite{Ju96} for a review, admits a realization on the real line in which the two supersymmetric generators can be assembled into a non hermitian supersymmetric charge of the form $Q=\frac{1}{2}\left(\sigma_1-i\sigma_2\right) D_y$, with $\sigma_k$ the Pauli matrices. The Hamiltonian is then a $2\times 2$ diagonal matrix $H=\frac{1}{2}\left\{ Q,Q^\dagger\right\}$ whose upper element is, apart of the factor $1\over 2$, precisely the Hessian ${\cal H}_y$. In the context of supersymmetric quantum mechanics, the existence of a normalizable zero mode of the Hessian, which was used in previous sections to solve the gravitational equations to linear order in $g$, is tantamount to the statement that supersymmetry is not spontaneously broken, while the standard kink solutions have the status of BPS object preserving half of the original $N=2$ supersymmetry. Models of supersymmetric quantum mechanics have important applications in physics \cite{Ju96}, and very remarkable ones in mathematics, as are the proof given by Witten of Morse inequalities in \cite{Wi82} or the physicist's proof of the Atiyah-Singer index theorem presented in \cite{AG83},\cite{FrWi84}.

All other members of the P\"{o}schl-Teller hierarchy enjoy, like the $\phi^4$ and sine-Gordon models, unbroken supersymmetry. Thus, in order to describe them, let us begin by formulating the perturbative approach to JT gravity for a model with this property. In general, the static limit of the field equations (\ref{eqjt1gauge})-(\ref{eqjt2gauge}) can be reverted to non-dimensional variables $y$ and $\psi$ by means of some convenient parameters $a$ and $b$ with mass dimensions one and zero through a rescaling
\beq
x=\frac{y}{a}\hspace{1.5cm} \phi=b\psi\hspace{1.5cm}V(\phi)=a^2 b^2 {\cal U}(\psi).\label{reparam}
\eeq
This makes ${\cal U}(\psi)$ also non-dimensional, and gives to the equations the form
\beqr
\frac{d}{dy}\left(\alpha\frac{d\psi}{dy}\right)&=&\frac{d{\cal U}}{d\psi}\label{eqjty1}\\
\frac{d^2\alpha}{dy^2}&=&g {\cal U}(\psi)\label{eqjty2}
\eeqr
where $g=16\pi G b^2$. This reparametrization matches with the changes applied in previous sections: for the sine-Gordon model $a=\sqrt{\lambda}$, $b=\frac{1}{\gamma}$ and for the $\phi^4$ theory $a=v\sqrt{\frac{\lambda}{2}}$, $b=v$, although the coupling denoted $g$ in Section 2 is half the coupling $g$ in our current notation. Let us also assume that, as it was the case for the $\phi^4$ and sine-Gordon models, the potential ${\cal U}(\psi)$ is a semidefinite positive, even function of $\psi$, with at least two consecutive symmetric vacua $\pm\psi_{\rm vac}$ such that ${\cal U}(\pm\psi_{\rm vac})=0$. In these circumstances there is, for zero gravitational coupling, a kink $\psi_K(y)$ of finite energy which lives in Minkowski spacetime, is an odd function of the coordinate $y$, and satisfies the Bogomolny equation
\beq
\frac{d\psi_K}{dy}=\sqrt{2{\cal U}(\psi_k)}\hspace{2cm} \psi_k(\pm\infty)=\pm\psi_{\rm vac}.\label{boggen}
\eeq
In fact, since
\bdm
\frac{d^2\psi_k}{dy^2}=\frac{1}{\sqrt{2{\cal U}(\psi_k)}}\left.\frac{d{\cal U}}{d\psi}\right|_{\psi=\psi_K}\frac{d\psi_K}{dy}=\left.\frac{d{\cal U}}{d\psi}\right|_{\psi=\psi_K}
\edm
the functions $\alpha(y)=1$ and $\psi(y)=\psi_K(y)$ solve the equations (\ref{eqjty1})-(\ref{eqjty2}) for $g=0$. Expanding around the zero-coupling kink as we did in (\ref{expa})-(\ref{expb}),  we find the equations at linear order in $g$
\beqr
{\cal H}_y\varphi&=&\frac{d}{dy}\left(\beta\frac{d\psi_K}{dy}\right)\label{eqpertjt1}\\
\frac{d^2\beta}{dy^2}&=&{\cal U}(\psi_K)\label{eqpertjt2},
\eeqr
where the Hessian operator is
\bdm
{\cal H}_y=-\frac{d^2}{dy^2}+\left.\frac{d^2{\cal U}}{d\psi^2}\right|_{\psi=\psi_K}.
\edm
Now, let us assume that the supersymmetric factorization (\ref{factor}) is valid in this model and that supersymmetry is unbroken, with the superpotential $W(y)$ being such that $W(+\infty)>0$ and $W(-\infty)<0$, as it happens in the $\phi^4$ and sine-Gordon theories. In such a case, we can proceed backwards, i.e., we can reconstruct the kink and, to some extent, even the potential of the scalar field theory starting from the Hessian \cite{BoYu03, BaBe17}. The reason is translational invariance, because this symmetry implies that ${\cal H}_y$ has always a normalizable zero mode, the translational mode of the kink, given by $f_0=\frac{d\psi_K}{dy}$
which, by virtue of unbroken supersymmetry, satisfies $D_y f_0=0$. Thus, for a kink centered at the origin, the translational mode is determined by the superpotential as
\beq
f_0(y)=A \exp\left(-\int_0^y dz W(z)\right)\label{zeromode}
\eeq
and, once we know the zero mode, we can integrate it to recover the kink profile 
\beq
\psi_K(y)=\int_0^y f_0(z) dz,\label{kink}
\eeq
where the constant $A$ is determined in terms of the vacuum expectation value $\psi_{\rm vac}$ through the asymptotic condition
\beq
\int_0^\infty f_0(z) dz=\psi_{\rm vac}\label{condi}.
\eeq
Also, from the zero mode of the kink it is possible to work out the potential energy of the model. The Bogomolnyi equation (\ref{boggen}) is tantamount to
\bdm
{\cal U}(\psi_K(y))=\frac{1}{2} f_0^2(y)
\edm
and this, together with (\ref{kink}), makes it feasible, at least in principle, to solve for ${\cal U}(\psi)$ for field values in the interval $[-\psi_{\rm vac},\psi_{\rm vac}]$. Finally, the energy of the kink follows also directly from the zero mode: the reparametrization (\ref{reparam}) gives $E[\phi_K]=a b^2 E_{\rm norm}[\psi_K]$, with the normalized energy and energy density given by
\beq
E_{\rm norm}[\psi_K]=\int_0^\infty\left\{\left(\frac{d\psi_K}{dy}\right)^2+ 2 {\cal U}(\psi_K)\right\}=2\int_0^\infty dy f_0^2(y),\hspace{2cm}{\cal E}_{\rm norm}^{\psi_K}(y)=2 f_0^2(y).\label{enernorm}
\eeq
The Bogomolny equation, on the other hand, implies that the pressure ${\cal P}_{\rm norm}^{\psi_K}(y)$ of the kink with $g=0$ vanishes. 
\subsection{The P\"{o}schl-Teller hierarchy of reflectionless potentials}
The P\"{o}schl-Teller hierarchy, reviewed for instance in \cite{Ma15}, provides a concrete realization of the previous scheme. This hierarchy is built by choosing a sequence of superpotentials of the form $W_\ell(y)=\ell \tanh(y)$ with $\ell$ a natural number. Thus, in particular, the superpotentials with $\ell=1$ and $\ell=2$ are those associated to the sine-Gordon and $\phi^4$ kinks. The Hessian operators are given by
\beq
({\cal H}_y)_\ell=(D^\dagger_y)_\ell(D_y)_\ell=\frac{d^2}{d y^2}+U_\ell(y)\hspace{2cm}U_\ell(y)=\ell^2-\ell(\ell+1) {\rm sech}^2(y)\label{hesspt}
\eeq
and exhibit potential wells which are deeper for increasing $\ell$. Incidentally, the free Hamiltonian can be included in the series by extending it to the case $\ell=0$. Each member in the sequence is related to the previous one by an interchange of the order of the first order differential operators, namely $(D_y)_\ell(D^\dagger_y)_\ell=({\cal H}_y)_{\ell-1}+\ell^2-(\ell-1)^2$. This property, called shape invariance, provides the basis of an algebraic method for solving the spectrum of $({\cal H}_y)_\ell$ which generalizes the factorization method originally developed by Schr\"{o}dinger, Infeld and Hull and others, see \cite{CaRa00}. Other remarkable features of the P\"{o}schl-Teller potentials are reflectionless scattering, the occurrence of a half-bound state at the threshold of continuous spectrum and the fact that the functional determinant can be computed exactly \cite{Ma15}. 

\begin{figure}
\centering
\includegraphics[width=7cm]{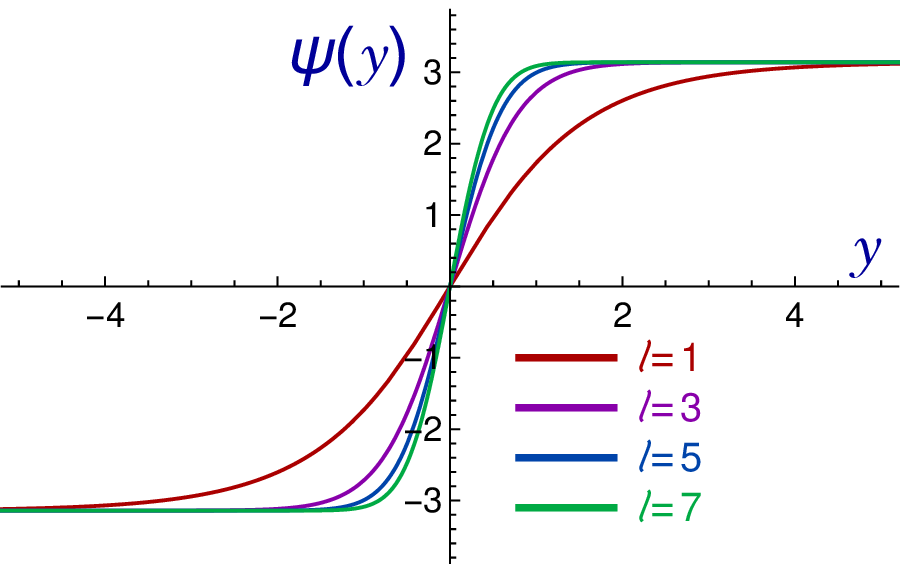}
\hspace{1.5cm}
\includegraphics[width=7cm]{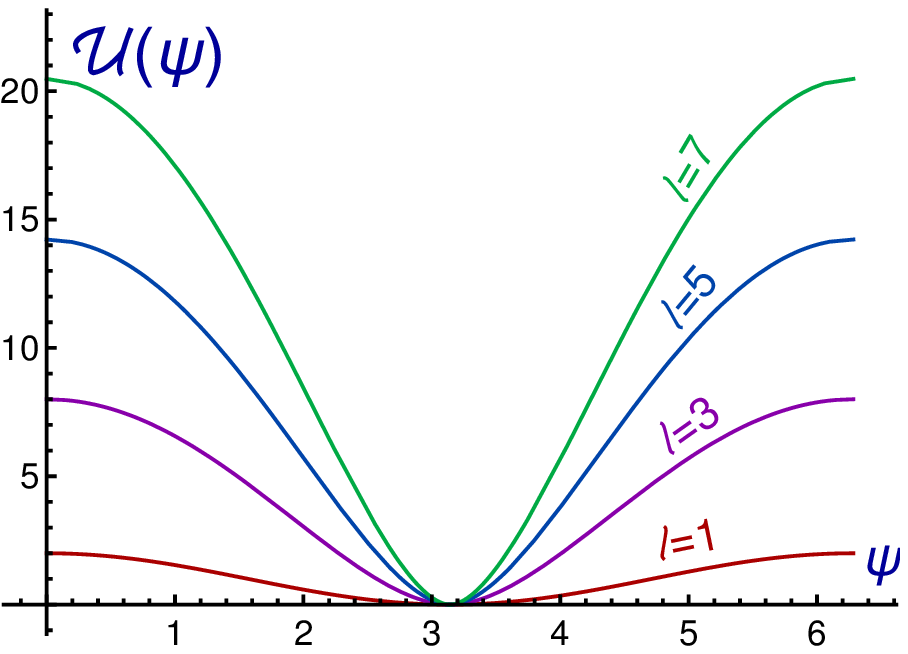}
\caption{The kink profile for zero gravitational coupling and the scalar potential for several odd values of $\ell$.}
\label{fig7}
\end{figure}

\begin{figure}
\centering
\includegraphics[width=7cm]{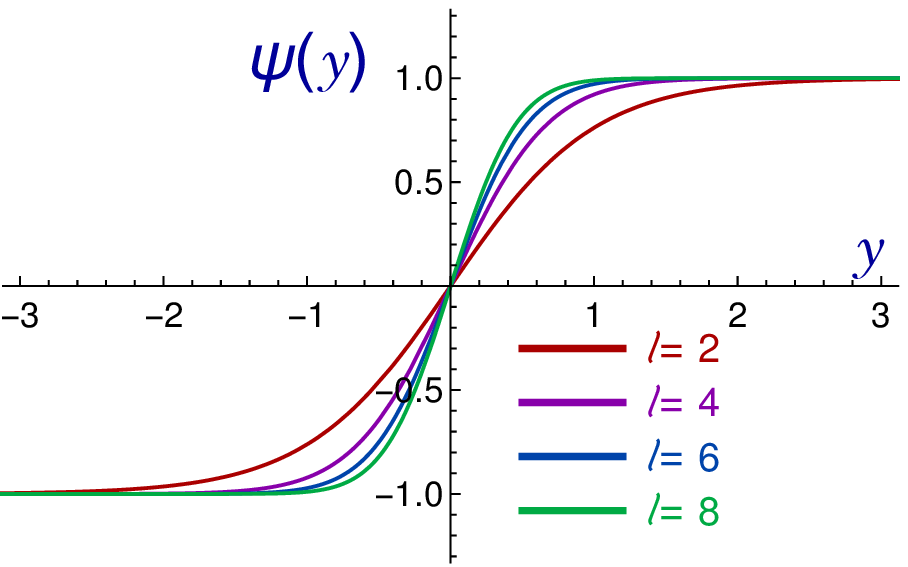}
\hspace{1.5cm}
\includegraphics[width=7cm]{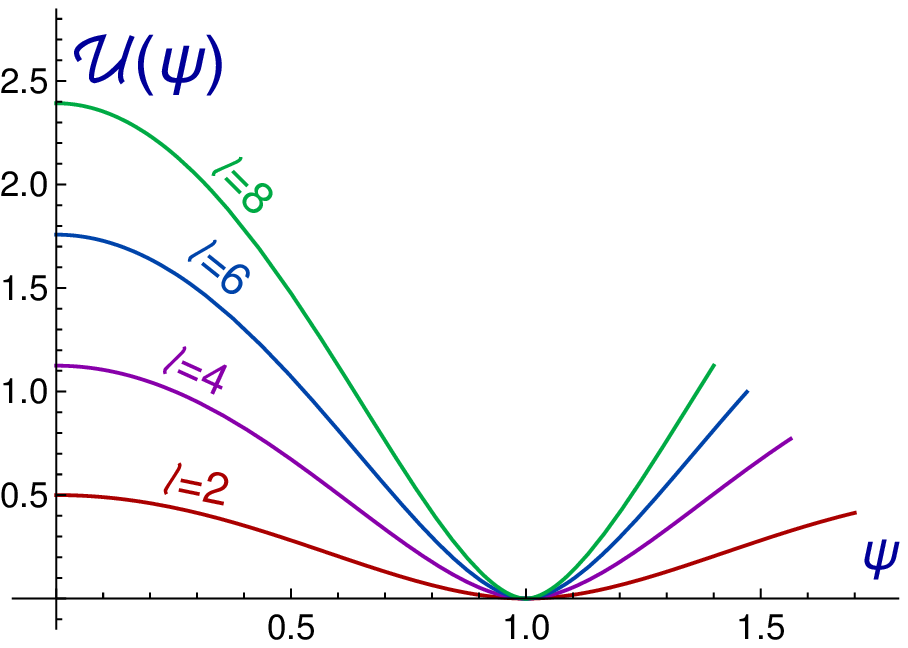}
\caption{The kink profile for zero gravitational coupling and the scalar potential for several even values of $\ell$.}
\label{fig8}
\end{figure}

As demanded by unbroken supersymmetry, all the models in the hierarchy have a normalizable zero mode, which using (\ref{zeromode}) turns out to be
\bdm
(f_0(y))_\ell=A_\ell\, {\rm sech}^\ell(y).
\edm
In order to match the usual conventions for the sine-Gordon and $\phi^4$ cases, it is convenient to choose different vacuum expectation values $\psi_{\rm vac}$ for the cases of even and odd $\ell$, according to
\bdm
\psi_{\rm vac}=\pi\  (\ell\ {\rm odd})\hspace{2cm}\psi_{\rm vac}=1\  (\ell\ {\rm even}),
\edm
and with this choice, the the $A_\ell$ factor coming from (\ref{condi}) is given by
\bdm
A_\ell=2 \pi^{\frac{(-1)^{\ell+1}}{2}}\frac{\Gamma(\frac{\ell+1}{2})}{\Gamma(\frac{\ell}{2})}.
\edm
On the other hand, we can use (\ref{kink}) to determine the profile of the kink leading to the Hessian $({\cal H}_y)_\ell$, and we arrive to the result
\bdm
(\psi_K(y))_\ell=A_\ell \int_0^y dz\;{\rm sech}^\ell(z)=A_\ell F\left(\frac{1}{2},\frac{\ell+1}{2};\frac{3}{2};-\sinh^2(y)\right)\sinh(y)
\edm
with $F$ the hypergeometric function. The next step is to figure out the scalar field theory which accommodates such a kink configuration. While in the cases of the sine-Gordon and $\phi^4$ models we have a our disposal an explicit expression of the potential ${\cal U}$ in terms of the field $\psi$, for higher $\ell$ the best that we can do is to proceed as was done in \cite{BoYu03} (see also \cite{AlJu12}) and to give both $\psi$ and ${\cal U}$ parametrically. For odd $\ell$, it is convenient to choose the parameter in the form $\tau={\rm sech}(y)$. As $\tau$ goes from $\tau=1$ to $\tau=0$, the field $\psi_K$ interpolates continuously between its values at the origin, $\psi_K(0)=0$ and at infinity, $\psi_K(\infty)=\pi$. The parametric expressions for the field and potential for odd $\ell$ are thus
\beqrn
\psi&=&A_\ell F\left(\frac{1}{2},\frac{\ell+1}{2};\frac{3}{2};\frac{\tau^2-1}{\tau^2}\right)\frac{\sqrt{1-\tau^2}}{\tau}\\
{\cal U}&=&\frac{1}{2} A_\ell^2 \tau^{2 \ell}
\eeqrn
and some particular cases are
\beqrn
\ell=1:&&\hspace{1cm} \psi=2 \arccos(\tau)\hspace{5cm} {\cal U}=2 \tau^2\\
\ell=3:&&\hspace{1cm} \psi=2t\sqrt{1-\tau^2}+2\arccos(\tau)\hspace{2.7cm} {\cal U}=8 \tau^6\\
\ell=5:&&\hspace{1cm} \psi=2t\sqrt{1-\tau^2}(1+\frac{2}{3} \tau^2)+2\arccos(\tau)\hspace{1.05cm} {\cal U}=\frac{128}{9} \tau^{10}.
\eeqrn
For even values of $\ell$ the explicit expressions are slightly simpler by defining the parameter as $\tau=\tanh(y)$. In this case, as $\tau$ interpolates between $\tau=0$ and $\tau=1$ the kink field varies from $\psi_K(0)=0$ to $\psi_K(\infty)=1$. The parametric expressions are
\beqrn
\psi&=&A_\ell F\left(\frac{1}{2},\frac{\ell+1}{2};\frac{3}{2};\frac{\tau^2}{\tau^2-1}\right)\frac{\tau}{\sqrt{1-\tau^2}}\\
{\cal U}&=&\frac{1}{2} A_\ell^2 (1-\tau^2)^\ell
\eeqrn
and some low-$\ell$ cases are
\beqrn
\ell=2:&&\hspace{1cm} \psi=\tau\hspace{4.5cm} {\cal U}=\frac{1}{2}(1-\tau^2)^2\\
\ell=4:&&\hspace{1cm} \psi=\frac{\tau}{2}(3-\tau^2)\hspace{3cm} {\cal U}=\frac{9}{8}(1-\tau^2)^4\\
\ell=6:&&\hspace{1cm} \psi=\frac{\tau}{8}(15-10 \tau^2+3\tau^4)\hspace{1.3cm} {\cal U}=\frac{225}{128}(1-\tau^2)^6.
\eeqrn
We show in figures \ref{fig7} and \ref{fig8} the kink profiles and the field theory potential for a few examples. Notice that the existence of the kink requires only that $\psi_{\rm vac}$ is a minimum of ${\cal U}$ with ${\cal U}(\psi_{\rm vac})=0$. Thus, as long as this condition is met, the potential  ${\cal U}(\psi)$ for $\psi>\psi_{\rm vac}$ can be chosen arbitrarily. Since it is not needed for our current purposes, we have made no attempt to fix this arbitrariness, and in the figures we simply have chosen ${\cal U}(\psi)$ symmetric around $\psi=\pi$ or $\psi=1$, at least near these points. Note, however, that well defined procedures to extend the potential beyond these limits, based on the single-valuedness of ${\cal U}(\psi)$ for complex values of the parametrization, have been developed \cite{BoYu03}, \cite{AlJu12}. Another interesting point is that if vacua with different absolute values of the vev are allowed, the reconstruction can result in a non-univocal answer, and different field theories can be recovered from the same Hessian \cite{BaBe17}.

Finally, we can use (\ref{enernorm}) to compute the normalized energy of the kink
\bdm
(E_{\rm norm}[\psi_K])_\ell=\ell\,\pi^{(-1)^{\ell+1}} \frac{2^\ell \Gamma^3\left(\frac{\ell+1}{2}\right)}{\Gamma(\ell+\frac{1}{2})\Gamma(\frac{\ell}{2}+1)}
\edm
and the normalized energy density, which has in all cases a maximum at the origin and decays with $y$ according to the expression
\bdm
({\cal E}_{\rm norm}^{\psi_K}(y))_\ell=2 A_\ell^2 {\rm sech^{2 \ell}}(y).
\edm
\subsection{Self gravitating P\"{o}schl-Teller kinks}
Having reviewed the main features of the kinks of and potentials of the P\"{o}schl-Teller hierarchy without gravity, we now come back to the case where $g$ is small but different from zero to look for solutions of the perturbative equations (\ref{eqpertjt1})-(\ref{eqpertjt2}) for these theories. Equation (\ref{eqpertjt2}), in conjunction with the usual boundary conditions (\ref{boundk1perta}) at the origin, can be integrated directly to give
\bdm
\beta_\ell(y)=\frac{A_\ell^2}{2}\int_0^y dz \int_0^z du\; {\rm sech}^{2\ell}(u)=\frac{A_\ell^2}{2}\int_0^y dz \int_0^z d \tanh(u)\, \left(1-\tanh^2(u)\right)^{(l-1)}.
\edm
Thus, the integral in $t$ leads to a sum of odd powers of hyperbolic tangents
\bdm
\int_0^z du\; {\rm sech}^{2\ell}(u)=\sum_{j=0}^{\ell-1} (-1)^j\left(\begin{array}{c}\ell-1\\j\end{array}\right)\frac{\tanh^{2j+1}(z)}{2j+1},
\edm
while the subsequent integration of these of powers gives rise to hypergeometric functions.
 All in all, the result for the perturbation of the metric is
\bdm
\beta_\ell(y)=\frac{A_\ell^2}{2}\sum_{j=0}^{\ell-1} (-1)^j\left(\begin{array}{c}\ell-1\\j\end{array}\right)\frac{\tanh^{2j+2}(y)}{(2j+1)(2j+2)} F\left(1,1+j;2+j;\tanh^2(y)\right).
\edm
Nevertheless, for later use of $\beta_\ell(y)$ into the integrals needed to compute the perturbation of the scalar field profile, it is more convenient to get rid of the hypergeometric functions. This can be done by means of formula 2.424-2 in \cite{GrRy80}, which gives an alternative expression for the integral of odd powers of the hyperbolic tangent as sums of even powers of hyperbolic cosines or tangents plus a logarithm. This formula, along with the identity 
\bdm
\sum_{j=0}^{\ell-1}\frac{(-1)^j}{2j+1}\left(\begin{array}{c}\ell-1\\j\end{array}\right)=\frac{\sqrt{\pi}\Gamma(\ell)}{2 \Gamma(\ell+\frac{1}{2})}
\edm
enables us to trade the previous expression for $\beta_\ell(y)$ for another one in which only elementary functions appear
\beq
\beta_\ell(y)=\frac{A_\ell^2}{2}\left\{\frac{\sqrt{\pi}\Gamma(\ell)}{2 \Gamma(\ell+\frac{1}{2})}\log(\cosh(y))-\sum_{i=1}^{\ell-1} C_{i,\ell} \tanh^{2i}(y)\right\},\label{altbeta}
\eeq
where the coefficients entering in the second term are
\bdm
C_{i,\ell}=\frac{1}{2i}\sum_{j=i}^{\ell-1} \frac{(-1)^j}{2j+1}\left(\begin{array}{c}\ell-1\\j\end{array}\right).
\edm

Let us now turn to the inhomogeneous Schr\"{o}dinger equation (\ref{eqpertjt1}). The potential well in the Hessian is given in (\ref{hesspt}), while the source
\bdm
R_\ell(y)=\frac{d}{dy}\left[ A_\ell \beta_\ell(y) {\rm sech}^\ell(y)\right]
\edm
can be written by means of the incomplete Euler beta function $B_z(a,b)=\int_0^z u^{a-1}(1-u)^{b-1} du$ as
\bdm
R_\ell(y)=\frac{A_\ell^3}{4}\sum_{j=0}^{\ell-1} (-1)^j\left(\begin{array}{c}\ell-1\\j\end{array}\right)\frac{2 \tanh^{2j}(y)-\ell B_{\tanh^2(y)}(j+1,0)}{1+2j}{\rm sech}^{\ell+1}(y) \sinh(y)
\edm
or, if one makes use of the version (\ref{altbeta}) of $\beta_\ell(y)$, as an expression with only hyperbolic functions 
\bdm
R_\ell(y)=\frac{A_\ell^3}{2}\left\{\frac{\sqrt{\pi}\Gamma(\ell)}{2 \Gamma(\ell+\frac{1}{2})}\left[1-\ell \log(\cosh(y))\right]\tanh(y)+\sum_{i=1}^{\ell-1} D_{i,\ell}(y) \tanh^{2i-1}(y){\rm sech}^2(y)\right\}{\rm sech}^\ell(y),
\edm
but with the occurrence of some awkward coefficients:
\bdm
D_{i,\ell}(y)=\frac{1}{2i}\sum_{j=i}^{\ell-1} \frac{(-1)^j}{2j+1}\left(\begin{array}{c}\ell-1\\j\end{array}\right)(\ell \sinh^2(y)-2 i).
\edm
Once the source is given in explicit form, we can solve the non homogeneous Schr\"{o}dinger equation by the procedure developed in Subsection 2.3. For the even normalizable zero mode we can take directly the translational mode
\bdm
\rho_\ell(y)=(f_0(y))_\ell={\rm sech}^\ell(y),
\edm 
while the odd non-normalizable one $\chi_\ell(y)$ is obtained by solving
\bdm
\frac{d\chi_\ell}{dy}+\ell\tanh(y) \chi_\ell=\cosh^\ell(y)
\edm
with the result
\beqrn
\chi_\ell(y)&=&{\rm sech}^\ell(y)\int_0^y dz \cosh^{2\ell}(z)=\frac{i y}{(2\ell+1)|y|}\cosh^{\ell+1}(y) F\left(\frac{1}{2},\frac{1}{2}+\ell;\frac{3}{2}+\ell;\cosh^2(y)\right)\\&-&\frac{i\sqrt{\pi}\Gamma(\ell+\frac{1}{2})}{2\Gamma(\ell+1)}{\rm \sech}^\ell(y),
\eeqrn
which can, once again, be more conveniently given as a sum of hyperbolic functions
\bdm
\chi_\ell(y)=\frac{1}{2^{2\ell}}\left[\left(\begin{array}{c}2\ell\\\ell\end{array}\right)y+\sum_{j=0}^{\ell-1}\left(\begin{array}{c}2\ell\\j\end{array}\right)\frac{\sinh(2(\ell-j)y)}{\ell-j}\right]{\rm sech}^\ell(y).
\edm
The normalization of zero modes has been chosen so that the Wronskian is unity. Now, to compute $\varphi_\ell(y)$ we have to do the integral (\ref{phi}). The integrand can be decomposed as a sum with terms made of products of powers of hyperbolic functions which, in some cases, also include a factor $\log(\cosh(z))$, or $z$, or both. Although it turns out that all these expressions can be integrated exactly, since both $\chi_\ell(y)$ and $R_\ell(y)$ involve sums with rather unhandy coefficients, to obtain a general expression valid for all $\ell$ appears to be quite cumbersome. Thus, here we content ourselves with giving the results for some low-$\ell$ members of the hierarchy, $3\leq\ell\leq 8$. In all these cases, the perturbation of the scalar field can be put in the form
\bdm
\varphi_\ell(y)=-M_\ell \left[\gamma(y)+ N_\ell\;y  - {\cal P}_\ell\left({\rm sech}^2(y)\right) \tanh(y)\right] {\rm sech}^\ell(y),
\edm
where $\gamma(y)$ is the function
\bdm
\gamma(y)=\frac{\pi^2}{12} + y (y - 2\log 2) + {\rm Li}_2(-e^{-2y}),
\edm
the factors $M_\ell$ and $N_\ell$ are rational numbers, and ${\cal P}_\ell\left(t\right) $ is a polynomial of degree $\ell-2$ with rational coefficients. The explicit values of the factors appear in the table
\begin{table}[H]
\renewcommand*{\arraystretch}{1.2}
\begin{center}
\begin{tabular}{||c|c|c|c|c|c|c||}
\hline
\multicolumn{7}{||c||}{\small Factors entering in $\varphi_\ell(y)$}\\
\hline
$\ell$&{\small 3}&{\small 4}&{\small 5}&{\small 6}&{\small 7}&{\small 8}\\
\hline
$M_\ell$&${{64\over 15}}$&${27\over140}$&${65536\over8505}$&${375\over1232}$&${4194304\over375375}$&${8575\over20592}$\\
\hline
$N_\ell$&${17\over48}$&${13\over24}$&${5069\over7680}$&${1901\over2560}$&${172889\over215040}$&${45791\over53760}$\\
\hline
\end{tabular}
\end{center}
\end{table}
\noindent and the polynomials are given in the following list:
\beqrn
{\cal P}_3(t)&=&{4\over 5} + {7 \over 80}t\\
{\cal P}_4(t)&=& {533\over630} + {31\over252} t+{5\over168} t^2\\  
{\cal P}_5(t)&=&{1745\over2016} + {569 \over 4032}t+ {35 \over 768}t^2 + {65 \over 4608}t^3 \\
\\ 
{\cal P}_6(t)&=& {45457\over 51975} + {2251 \over 14850}t+ {34907 \over 633600}t^2 + {641 \over 28160}t^3 + {7 \over 880}t^4\\     
{\cal P}_7(t)&=& {904757\over 1029600} + {325607 \over 2059200}t+ {168307 \over 2745600}t^2 + {93947 \over 3294720}t^3 + 
               {3983 \over 299520}t^4 + {133 \over 26624}t^5\\ 
{\cal P}_8(t)&=&  {1233833\over 1401400} + {4097497 \over 25225200}t + {2205607 \over 33633600}t^2 + {119437 \over 3669120}t^3 + 
               {35831 \over 2096640}t^4 +{307 \over 35840}t^5 + {121 \over 35840}t^6.        
\eeqrn

\begin{figure}[t]
\centering
\includegraphics[width=7cm]{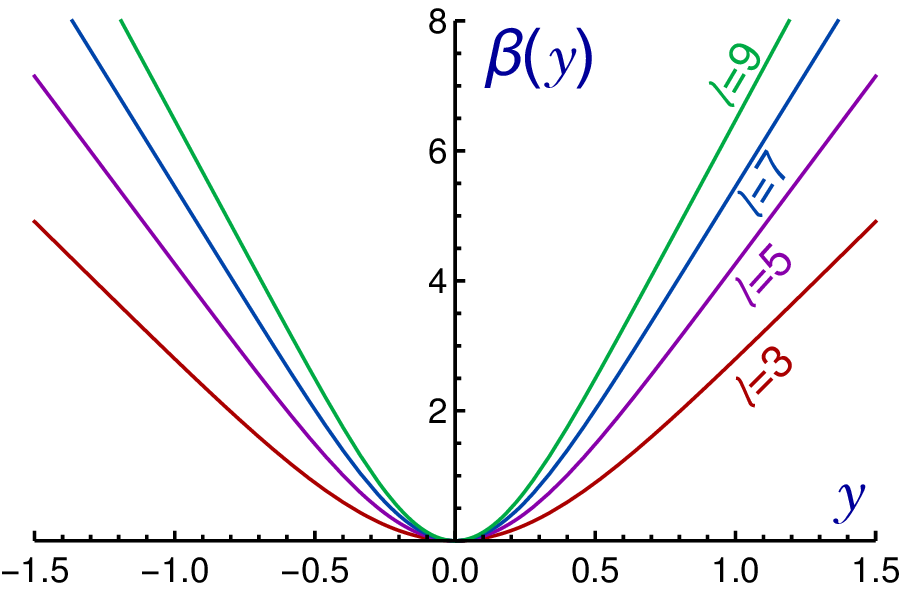}
\hspace{1cm}
\includegraphics[width=9cm]{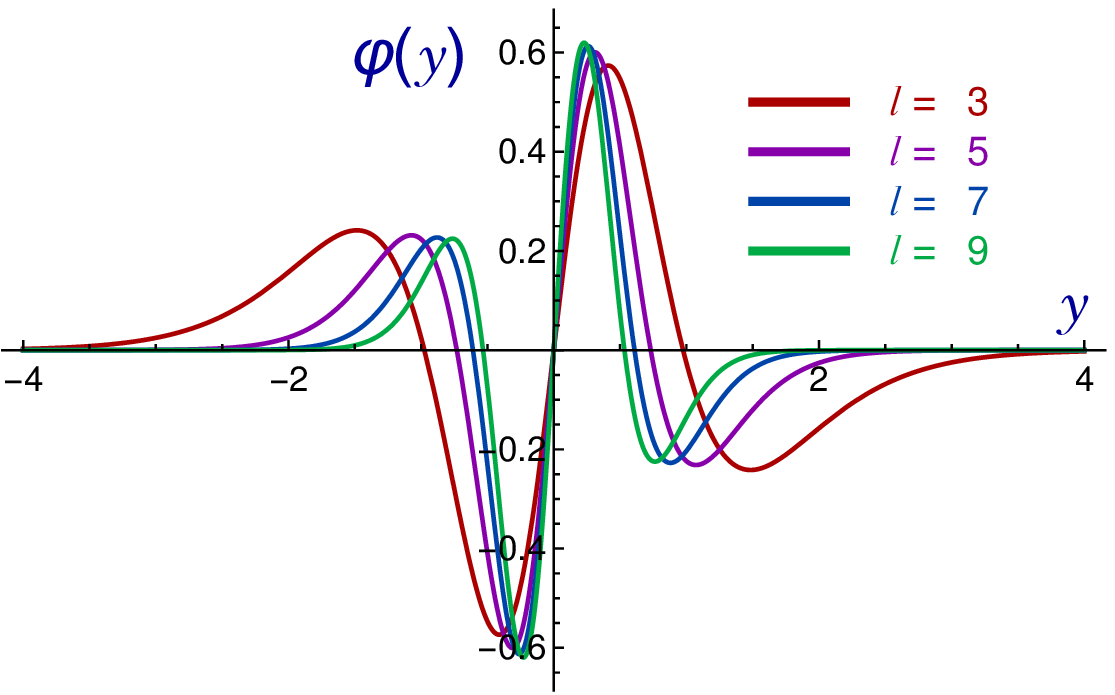}
\caption{The perturbations of the metric and scalar kink profile for several odd values of $\ell$.}
\label{fig9}
\end{figure}

\begin{figure}[b]
\centering
\includegraphics[width=7cm]{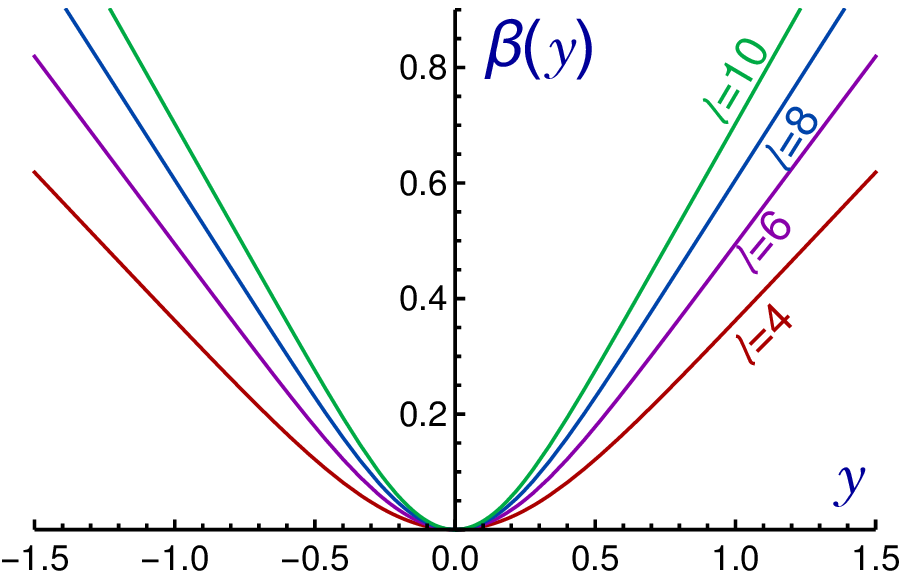}
\hspace{1cm}
\includegraphics[width=9cm]{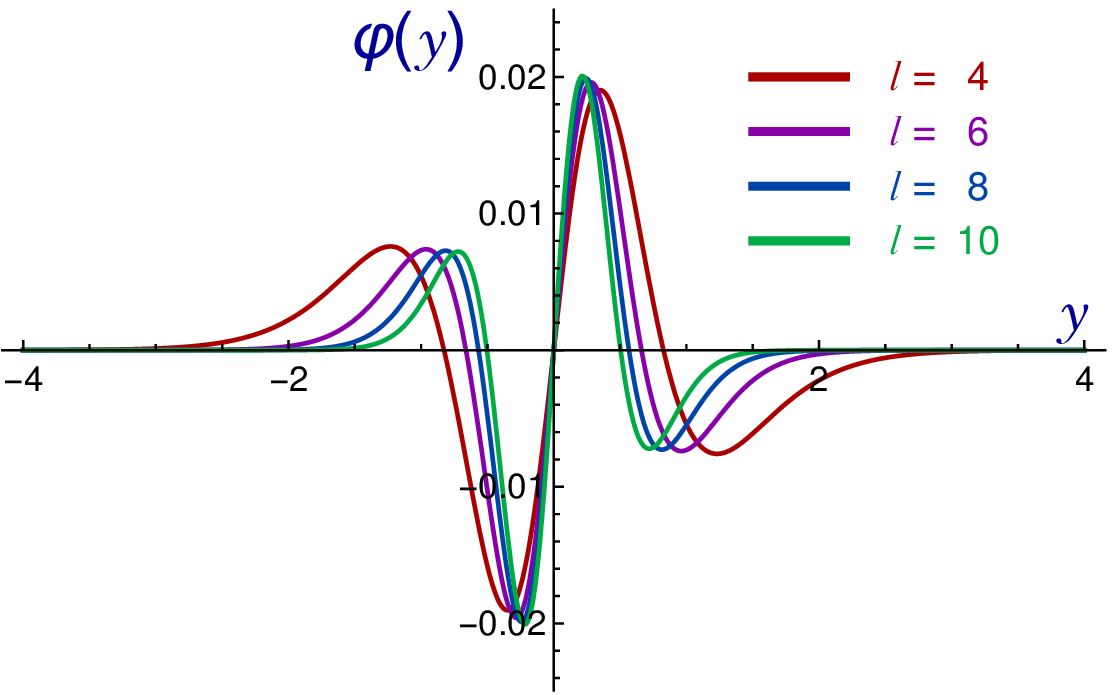}
\caption{The perturbations of the metric and scalar kink profile for several even values of $\ell$.}
\label{fig10}
\end{figure}

We present also some graphics in figures \ref{fig9} and \ref{fig10}. As one can see, the general features of the metric and scalar field perturbations are the same irrespective of the value of $\ell$, the differences being only quantitative. For each parity, the function $\beta_\ell(y)$ increases with $|y|$ at a rate which is rising as $\ell$ becomes higher. As for the function $\varphi_\ell(y)$, it displays a sort of damped oscillations around the origin before decaying for high $|y|$. For $y>0$, it first reaches a maximum which is higher and closer to the origin as $\ell$ increases, and then a minimum, this time shallower and also closer to the origin for increasing $\ell$.
\section{Outlook}
Along this paper we have been investigating the interplay between Jackiw-Teitelboim gravity and travelling kink solutions in severals scalar field theories. According to the classification given by Bazeia in \cite{Ba02}, all these theories are of type I, models with a single scalar field supporting structureless kinks. In fact, they all belong to a subclass in which the energy density is symmetric around the center of the kink, a condition which does not apply for other models within type I such as the $\phi^6$ kink \cite{Lo79}. There is also a second type of models which, although contain also a sole field, are able to embrace kinks of two different species, for instance the double sine-Gordon model introduced in \cite{MaKu76}. Finally, type III comprises a variety of models with several scalar fields and where the interactions among different kink components lend them internal structure. In models of this type, when the potential has non collinear minima, it is possible to engineer junction configurations of kinks, see some examples in \cite{Ba02}, or to consider a non flat target manifold to uncover the presence of kinks in nonlinear sigma models \cite{Juanetal1,Juanetal2}. It would be interesting to put JT gravity into models of types II and III and to explore the consequences of gravitational physics on the rich dynamics enjoyed by these theories.

The scope of the treatment that we have given here is limited to the presentation of the self-gravitating kink solutions arising in the models considered, but without a detailed analysis of their stability or possible quantization, which can be the subject of further work. For zero gravitational coupling, the stability of kinks is very solid due to topological reasons. Since, in presence of gravity, finiteness of energy imposes the same constraints on the asymptotic behavior of the scalar field than in the non gravitational case, we expect the self-gravitating kinks to be stable against scalar field fluctuations, at least for small $g$ coupling. However, fluctuations of the metric fields have also a role, and they could give rise to instability modes leading to gravitational collapse into a black hole with topologically non-trivial boundary conditions. This could be specially significant when $g$ is higher and the kink energy increases. A criterion for gravitational stability for pressureless matter has been worked out in \cite{SiMa91}, but the situation here is more complicated and further study is required to elucidate this point. In any case, the computation of the spectrum of kink fluctuations and the study of scalar and gravitational waves in a kink background are interesting issues which deserve a careful examination.

As we have mentioned in the Introduction, the theory of Jackiw and Teitelboim is only a particular case, although a fairly interesting one,  within the category of two-dimensional dilaton gravity theories, which can be formulated in quite a variety of ways. The presence of black holes in these theories and the diverse gravitational phenomenology springing up in them has been a theme of considerable research \cite{GrKuVa02}. Thus, to deal with other dilaton theories is, along with the consideration of the wide diversity of scalar models alluded above, another natural direction in which the results reported in the present work can be extended. Finally, although taking the cosmological constant equal to zero, as we have done, is the most natural option in the prospect for obtaining kink solutions, taking into account the effect of a non vanishing $\Lambda$ on field configurations like those described in the paper could be another problem to be addressed.
\section*{Acknowledgments}
The authors acknowledge the Junta de Castilla y Le\'on for financial help under grants SA067G19 and BU229P18.

\end{document}